\documentclass[utf8]{frontiersSCNS} 

\usepackage{url,hyperref,lineno,microtype,caption,subcaption,multirow}
\usepackage[onehalfspacing]{setspace}

\def\keyFont{\fontsize{8}{11}\helveticabold }
\def\firstAuthorLast{Raufi {et~al.}} 
\def\Authors{Bujar Raufi\,$^{1,*}$, Luca Longo\,$^{1}$}
\def\Address{$^{1}$Artificial Intelligence and Cognitive Load Lab, The Applied Intelligence Research Centre, School of Computer Science, Technological University Dublin, Republic of Ireland}
\def\corrAuthor{Bujar Raufi, Luca Longo}

\def\corrEmail{bujar.raufi@tudublin.ie, luca.longo@tudublin.ie}

\newcommand{\varB}[1]{{\operatorname{\mathit{#1}}}}
\begin{document}
\onecolumn
\firstpage{1}  

\title[]{An Evaluation of the EEG alpha-to-theta and theta-to-alpha band Ratios as Indexes of Mental Workload} 

\author[\firstAuthorLast ]{\Authors} 
\address{} 
\correspondance{} 

\extraAuth{}

\maketitle
\begin{abstract}

\section{}
Many research works indicate that EEG bands, specifically the alpha and theta bands, have been potentially helpful cognitive load indicators. However, minimal research exists to validate this claim. This study aims to assess and analyze the impact of the alpha-to-theta and the theta-to-alpha band ratios on supporting the creation of models capable of discriminating self-reported perceptions of mental workload. A dataset of raw EEG data was utilized in which 48 subjects performed a resting activity and an induced task demanding exercise in the form of a multitasking SIMKAP test. Band ratios were devised from frontal and parietal electrode clusters. Building and model testing was done with high-level independent features from the frequency and temporal domains extracted from the computed ratios over time. Target features for model training were extracted from the subjective ratings collected after resting and task demand activities. Models were built by employing Logistic Regression, Support Vector Machines and Decision Trees and were evaluated with performance measures including accuracy, recall, precision and f1-score. The results indicate high classification accuracy of those models trained with the high-level features extracted from the alpha-to-theta ratios  and theta-to-alpha ratios. Preliminary results also show that models trained with logistic regression and support vector machines can accurately classify self-reported perceptions of mental workload. This research contributes to the body of knowledge by demonstrating the richness of the information in the temporal, spectral and statistical domains extracted from the alpha-to-theta and theta-to-alpha EEG band ratios for the discrimination of self-reported perceptions of mental workload.  

\tiny
 \keyFont{ \section{Keywords:} human mental workload, EEG band ratios, alpha-to-theta ratios, theta-to-alpha ratios, machine learning, classification} 
\end{abstract}

\section{Introduction}
\label{sec:intro}
Human mental workload is a fundamental concept for investigating human performance. It represents an intrinsically complex and multilevel concept, and ambiguities exist in its definition. The most general description of mental workload can be framed as the quantification of a cognitive cost of performing a task in a finite timeframe in order to predict operator, system performance or both \citep{hancock2021mental, RizzoL18Inferential, reid_subjective_1988}.
Mental workload has been regarded as an essential factor that substantially influences task performance \citep{young_state_2015, galy2018consideration, longo2018experienced}. As a construct, it has been widely applied in the design and evaluation of complex human-machine systems and environments such as in aircraft operation \citep{yu2021correction, hu2020detecting}, train and vehicle operation\citep{wang2021mental, li2020exploring}, nuclear power plants \citep{gan2020workload, wu2020using}, various human-computer and brain-computer interfaces \citep{asgher2020enhanced, longo2012formalising, bagheri2021investigating, putze2020brain} and in educational contexts \citep{Moustafa2019, OrruL19, Longo2020Evaluating, Longo2021}, to name a few. Mental workload research has accumulated momentum over the last two decades, given the fact that numerous technologies have emerged that engage users in multiple cognitive levels and requirements for different task activities operating in diverse environmental conditions.

Different methods have been proposed to measure human mental workload. These methods can be clustered into three main groups. \emph{Subjective measures} which relies on the analysis of the subjective feedback provided by humans interacting with an underlying task and is usually in the form of a post-task survey. The most well-known subjective measurement techniques are the NASA Task Load Index (NASATLX) \citep{hart1988development}, the Workload profile (WP) \citep{tsang1996diagnosticity}, and the Subjective Workload Assessment Technique (SWAT) \citep{reid_subjective_1988}. \emph{Task performance measures}, often referred to as primary and secondary tasks measures, focus on the objective measurement of a human's performance in an underlying task. Examples of such measures include timely completion of a task, reaction time to secondary tasks, number of errors on the primary task and tapping error. \emph{Physiological measures} are based upon the analysis of physiological responses of the human body. Examples include EEG (electroencephalography), MEG (magnetoencephalography), brain metabolism, endogenous eye blink rates, pupil diameter, heart rate variability (HRV) measures or electrodermal responses such as galvanic skin response (GSR) \citep{byrne2011measurement}.

Many research works indicate that EEG data contains information that can help correlate task engagement and mental workload in cognitive processes like vigilance, learning and memory \citep{berka2007eeg, roy_efficient_2016}, in operating under environmental factors such as temperature \citep{wang_investigating_2019} and in critical systems domains such as transport \citep{diaz2020eeg, borghini2014measuring}, nuclear power plants \citep{choi2018development} and aviation \citep{wilson2021identifying}. The reason for using EEG is that it offers several benefits compared to imaging techniques or mere behavioural observational approaches. The most important benefit of EEG is its excellent time resolution which offers the possibility to study the precise time-course of cognitive and emotional processing of behaviour. Billions of neurons in the human brain are organized in a highly intricate and convoluted fashion exhorting in complex firing patterns. These patterns, accompanied by frequency oscillations, are measurable with EEG reflecting certain cognitive, affective or attentional states. These frequencies, in adults, are usually decomposed in different bands: delta band (1 – 4 Hz), theta band (4 – 8 Hz), the alpha band (8 – 12 Hz), the beta band (13 – 25 Hz) and gamma band ($\ge$ 25 Hz) \citep{mesulam1990report}.

Recent studies seem to indicate changes in frequency band across different brain regions when a subject performs specific tasks \citep{7905130, gevins2003neurophysiological, 10.3389/fnagi.2013.00060}. The theta band is thought to be linked to mental fatigue and mental workload \citep{gevins1995towards}. The increase in theta spectral power is thought to be correlated with the rise in the use of cognitive resources \citep{xie2016effects, tsang2006mental}, task difficulty \citep{antonenko2010using} and working memory\citep{borghini2012assessment}. Alpha band tends to show sensitivity in experiments  with mental workload \citep{xie2016effects, puma2018using}, cognitive fatigue \citep{borghini2012assessment}, attention and alertness \citep{kamzanova2014use}. 

Even though EEG bands have been proposed as indicators that can discriminate mental workload \citep{antonenko2010using, tsang2006mental, gevins2003neurophysiological, coelli2015eeg}, it is unclear which of these best contribute to such discrimination. This article aims to identify the impact of the high-level features extracted from alpha and theta band ratios (and their combination) on the discrimination of levels of perception of mental workload self-reported by users. To tackle this aim, an empirical research experiment has been designed to generate time-series of alpha, theta band ratios, and their combinations, and extract high-level features that can be used to build models to classify self-report perceptions of mental workload.

The remainder of this article is organized as follows: section \ref{sec:relatedwork} outlines the related work regarding the specific definition and use of the alpha-to-theta and theta-to-alpha band ratios along with their relationship to mental workload. Section \ref{sec:design} describes the design of an empirical experiment and the methodology employed for answering the above research goal. Section \ref{sec:results} presents the findings followed by a critical discussion while section \ref{sec:conclusions} concludes this work, proposing future research directions.

\section{Related Work}
\label{sec:relatedwork}
Recent studies analyze EEG bands on various experimental settings designed for specific domains and purposes such as fatigue and drowsiness \citep{borghini2014measuring}, brain-computer interfaces \citep{kathner2014effects, gevins2003neurophysiological}, learning \citep{7905130, dan2017real} as well as for specific brain function disorders such as Alzheimer \citep{10.3389/fnagi.2013.00060}. Most research studies seem to indicate the possibility that EEG signals across various cortical regions can be a helpful tool towards discriminating mental workload while performing experiments with varying degree of task demands \citep{borghini2014measuring}. 

The theta band is thought to be linked to mental fatigue and drowsiness \citep{borghini2014measuring, gevins1995towards}. Increase of spectral power in the theta band is associated with an increase of demand in cognitive resources \citep{xie2016effects, tsang2006mental}, an increase in task difficulty \citep{antonenko2010using, kathner2014effects,gevins2003neurophysiological,borghini2015avionic} and an increase in working memory \citep{borghini2012assessment,borghini2014measuring}. Particularly, the theta power spectrum seems to increase in cases where a prolonged concentration while executing a task is required \citep{gevins2003neurophysiological, borghini2014measuring, kathner2014effects}. Some research even indicates a decrease in vigilance and alertness where a higher power spectrum in theta band is observed \citep{kamzanova2014use}. The brain regions thought to be associated with theta activity are mostly in the frontal cortical area \citep{gevins2003neurophysiological, borghini2014measuring, dan2017real}.

The alpha band seems to indicate sensitivity towards mental workload \citep{xie2016effects,puma2018using}, cognitive fatigue \citep{borghini2012assessment, borghini2014measuring}, and an increase in the alpha band activity is associated with a decrease in attention and alertness \citep{kamzanova2014use}. An increase/decrease in the alpha band power spectrum is witnessed during relaxed states with eyes closed and opened, respectively \citep{antonenko2010using}. A continuous suppression in the alpha band seems to be linked with increments of task difficulty \citep{mazher2017eeg}. The brain regions that are primarily associated with the alpha-band activity are parietal and occipital areas \citep{borghini2014measuring, puma2018using}. 

The beta band is thought to be linked to visual attention \citep{wrobel2000beta}, short term memory tasks \citep{palva2011localization} and inconclusively it was hypothesized that an increase in the beta band is associated with an increase in working memory \citep{spitzer2017beyond}. An increase in the beta band spectrum seems to be associated with increased levels of task engagement \citep{coelli2015eeg} and concentration \citep{kakkos2019mental}. The brain regions that are associated with the beta-band activity are parieto-occipital areas that have been observed during visual working memory task experiments \citep{mapelli2019brain}.

Multiple EEG band combinations and ratios have also been used to improve mental workload assessment. For instance, $beta / (alpha + theta)$ known as engagement index is used to study task human engagement \citep{mikulka2002effects}, mental attention \citep{maclean2012resting} and mental effort \citep{smit2005mental}. The reduction in the alpha band activity seems to correlate with increased activity in the frontal-parietal areas with an increase in beta power followed by a decrease in theta, which indicates high vigilance states \citep{maclean2012resting}. Alpha band activity reduction is also thought to correlate with activities in the parietal brain region where a decrease in beta activity followed by an increase of theta band activity indicate states of drowsiness and low attention \citep{maclean2012resting}.

Attempts to assess mental workload and task engagement using the information from the theta and alpha bands in the form of theta-to-alpha band ratios are seen in \citep{gevins2003neurophysiological, kathner2014effects,dan2017real,xie2016effects}. This is based on the assumption that an increase in the theta power band in the frontal brain region, and a decrease in the alpha power in parietal region is associated with an increase in mental workload \citep{kathner2014effects}. The increase in both alpha and theta power is related to the rise of fatigue \citep{kathner2014effects,xie2016effects}. Research seems to indicate that task load manipulations are followed by an increase of theta band activity in frontal brain regions followed by a decrease in alpha power in the parietal areas \citep{kathner2014effects, dan2017real,gevins2003neurophysiological}.

The motivation for this article arises from the fact that research studies are indicating that band ratios, specifically the theta and alpha bands, are associated with mental workload states \citep{gevins2003neurophysiological, borghini2014measuring} and to some extent, this seems to justify their potential as workload indicators \citep{fernandez2020electroencephalographic}. While research exists that focuses on the alpha, theta and beta bands as well as their respective ratios such as $beta / (alpha + theta)$ and to some extent, theta-to-alpha, there is an absence of research related to the use of the alpha-to-theta and theta-to-alpha ratios and their role in discriminating self-reported perceptions of mental workload. Therefore, to address the goal as stated in the introductory section \ref{sec:intro} we formulate a research problem focused on the investigation of the importance of high-level features extracted from the alpha-to-theta and the theta-to-alpha EEG band ratios on the discrimination of levels of perception of mental workload. In other words, the research question that can be formulated is: what is the impact of high-level features extracted from alpha and theta band ratios (and their combination) on discriminating of levels of perception of mental workload self-reported by users?

\section{Design and Methodology}
\label{sec:design}
\begin{figure}[!ht]
\centering
\includegraphics[width=18cm]{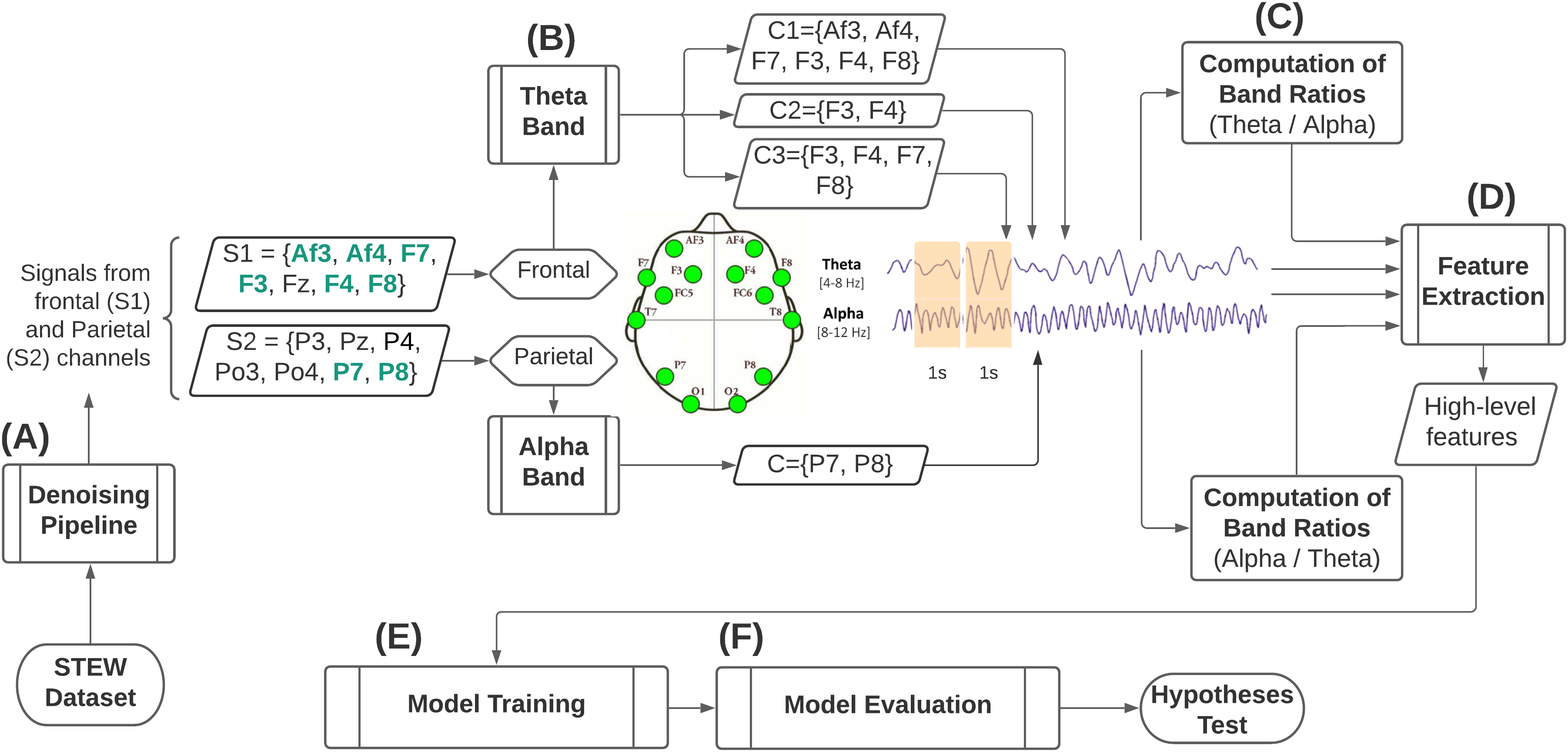}
\caption{Illustrative process for classification of self-reported perception of mental workload based on mental workload indexes built upon the EEG alpha and theta bands. \textbf{A)} Signal denoising pipeline. \textbf{B)} Electrode selection for theta band from frontal cortical areas and alpha band from parietal cortical areas and their aggregation to form electrode clusters. \textbf{C)} Computation of the mental workload indexes employing the EEG alpha-to-theta and theta-to-alpha band ratios. \textbf{D)} Extraction of high level features from mental workload indexes. \textbf{E)} Model training for self-reported perception of mental workload classification employing machine learning. \textbf{F)}. Model evaluation for hypothesis testing.}
\label{fig:1}
\end{figure}

To answer the research problem and research question outlined above, the following research hypotheses are defined:
\begin{enumerate}
    \item \textbf{H1:} If high-level features are extracted from indexes of mental workload built upon alpha-to-theta and theta-to-alpha band ratios, then their discriminatory capacity to self-reported perceptions of mental workload will be higher than those extracted from indexes of mental workload built upon the alpha and theta bands alone.
    \item \textbf{H2:} If more adjacent EEG electrodes from the respective cortical areas are used to create indexes of mental workload built upon alpha-to-theta and theta-to-alpha band ratios, then they will exhibit higher discriminatory capacity to self-reported perceptions of mental workload than those indexes built with fewer electrodes.
\end{enumerate}
In order to test these research hypotheses, empirical comparative research has been designed based on a  process pipeline as illustrated in figure \ref{fig:1} with details outlined in the following subsections.
\subsection*{Experiment Design and Dataset Description}
The STEW (Simultaneous Task EEG Workload) \citep{lim2018stew} has been selected for experimental purposes. The dataset consists of raw EEG data collected from 48 subjects across 14 channels in two experimental conditions. In one condition, the EEG data was recorded from subjects in the rest state while not performing any mental activity. In the second condition, a multitasking SIMKAP test was presented to subjects, and EEG data was recorded. In both cases, a sampling frequency of 128Hz was used with 2.5 minutes of EEG recordings utilizing the Emotiv EPOC EEG headset. Every recording contains 19,200 data samples (128 samples x 150 seconds) across the following 14 channels: AF3, F7, F3, FC5, T7, P7, O1, O2, P8, T8, FC6, F4, F8, AF4. Additionally, a subjective rating was collected after each task execution whereby users rated their experienced mental workload on the scale 1 to 9. The rating was a likert scale with 1 = ``very, very low mental effort''; 2 = ``very low mental effort''; 3 = ``low mental effort''; 4 = ``rather low mental effort''; 5 = ``neither low nor high mental effort''; 6 = ``rather high mental effort''; 7 = ``high mental effort''; 8 = ``very high mental effort'' and 9 = ``very, very high mental effort''. The rationale of using a perceived mental workload scores, was  a form of subjective validation to verify whether a subject indeed experienced an increase in cognitive load load while performing the SIMKAP condition as compared to the resting condition.

\renewcommand\thesubsection{\Alph{subsection}}
\subsection{EEG Denoising Pipeline}
Applying a denoising pipeline is an important step to pre-process the raw EEG data and to remove noise from it to facilitate subsequent analysis. In detail, this process follows the Makoto's pre-processing pipeline \citep{miyakoshi2018makoto} including:
\begin{itemize}
\item re-referencing channel data to average reference
\item high-pass filtering of each channel at 1hz
\item source separation and artefact removal via Independent Component Analysis (ICA)
\end{itemize}
The key pre-processing step is the application of ICA which is utilised to separate the 14 EEG signal sources into independent components for each subject. Fourteen components are generated and used to automatically find and remove artefacts without human intervention using part of the methodology described in \citep{NOLAN2010152}. In detail, the criteria for identifying bad components includes the computation of the z-scores of each component's spectral kurtosis, slope, Hurst exponent and gradient median. Spectral kurtosis is a parameter in the frequency domain indicating the component's impulsiveness variation with frequency. The slope of a component represents its mean slope of the power spectrum over two-time points. The Hurst exponent, also known as a long term memory in time series, tends to measure the tendency of a component to either regress to its mean or to catch up with an upward/downward trend. The gradient median is the median slope of the component's time course. All of the components exhorting values above and below ranges ``$\varB{z-score}\pm 3$'' can be considered as artefacts since they are outliers and significantly different from all the others. Finally, the inverse ICA has been executed to convert the remaining ``good'' components back in the original neural EEG signal.

\subsection{Forming Cluster Combinations}
A baseline of initial parietal and frontal electrodes was adopted following the electrode locations from the 10-20 international system to form different alpha and theta clusters for analysis and comparison purposes. These electrode locations were cross-referenced with locations, naming notation and electrode availability from the Emotiv EPOC EEG headset. The initial electrodes that are selected from the frontal and parietal locations  are indicated as $\varB{S1}$ and $\varB{S2}$ in figure \ref{fig:1}. Due to the limited availability of electrodes from the Emotiv EPOC EEG headset (highlighted in green in figure \ref{fig:1}), three frontal and one parietal clusters were constructed. In detail, cluster combinations and electrodes, together with the channel aggregation approach used, is shown in table \ref{tab:table-1}.
\begin{table}[!ht]
\centering
\caption{\label{tab:table-1}Clusters and electrode combinations from frontal and parietal cortical regions selected from the available electrodes.}
\renewcommand{\arraystretch}{1.1}%
\begin{tabular}{c c c c p{7cm}} 
\hline
 Cluster notation & Band & Electrodes & Channel aggregation approach \\ [0.5ex]
\hline\hline
 $\varB{c1-\theta}$ & theta & AF3, AF4, F3, F4, F7, F8 &  Average  \\ 
 $\varB{c2-\theta}$ & theta & F3, F4 & Average\\
 $\varB{c3-\theta}$ & theta & F3, F4, F7, F8 & Average \\
 $\varB{c-\alpha}$ & alpha & P7, P8 & Average\\
 \hline
\end{tabular}
\end{table}

\subsection{Formation of the Mental Workload indexes from clusters of EEG alpha and theta bands}
Generating band ratios from EEG channels over time follows the methodology used in \cite{borghini2014measuring}. The computation of the alpha-to-theta and theta-to-alpha ratios was done utilizing the average power spectral density (PSD) values from the alpha band from the cluster $\varB{c-\alpha}$ and the average PSD values from the theta band from clusters $\varB{c1-\theta}$, $\varB{c2-\theta}$ and $\varB{c3-\theta}$ as outlined in table \ref{tab:table-1}. Three alpha-to-theta and theta-to-alpha ratios are set to take different clusters from frontal electrodes and one cluster from parietal electrodes. The computation of the band ratios are given as follows:
\begin{equation}
\dfrac{\alpha}{\theta}  = \frac{avg (\forall e\in \ \varB{c-\alpha}) }{avg( \forall e\in \ \varB{c_x-\theta} )}
\end{equation}
\begin{equation}
\dfrac{\theta}{\alpha}  = \frac{avg (\forall e\in \ \varB{c_x-\theta}) }{avg(\forall e\in \ \varB{c-\alpha})}\end{equation}
where, $\varB{c-\alpha}$ and $\varB{c_x-\theta}$ are the respective alpha and theta clusters (from table \ref{tab:table-1}), with $e$ an electrode in them, and $x$ a cluster among those using the theta band ($\varB{c1-\theta}$, $\varB{c2-\theta}$, $\varB{c3-\theta}$). The  combination of the clusters in table \ref{tab:table-1}, jointly with their individual use, led to the formation of the following possible mental workload indexes (configurations):
\begin{equation}
\label{eq:eq-3}
   \varB{MWL\;Indexes}=\{\varB{c1-\theta}, \varB{c2-\theta}, \varB{c3-\theta}, \varB{c-\alpha}, \varB{at-1}, \varB{at-2}, \varB{at-3}, \varB{ta-1}, \varB{ta-2}, \varB{ta-3}\} 
\end{equation}
where, $\varB{at-1}=\dfrac{\varB{c-\alpha}}{\varB{c1-\theta}}$, $\varB{at-2}=\dfrac{\varB{c-\alpha}}{\varB{c2-\theta}}$, $\varB{at-3}=\dfrac{\varB{c-\alpha}}{\varB{c3-\theta}}$, $\varB{ta-1}=\dfrac{\varB{c1-\theta}}{\varB{c-\alpha}}$, $\varB{ta-2}=\dfrac{\varB{c2-\theta}}{\varB{c-\alpha}}$ and $\varB{ta-3}=\dfrac{\varB{c3-\theta}}{\varB{c-\alpha}}$.
In this study, a 1 second non-overlapping sliding widow technique is employed to segment long EEG data, and for each window, an index of mental workload can be calculated.

\subsection{Feature Extraction from indexes and Selection}
Extracting high-level features from MWL indexes is crucial since it allows the finding of distinguishing properties that otherwise would not be possible if a raw index alone is considered. The extraction of such high-level features from the indexes defined in the set \ref{eq:eq-3} is executed using the TSFEL (Time Series feature Extraction Library) \citep{barandas2020tsfel}. The advantage of using TSFEL is that it offers a wide range of statistical properties that can be extracted from multiple domains, including those from frequency and temporal domains. It is useful for identifying peculiar aspects of a signal and its specific properties such as variability, slope or peak to peak, just to name a few. Classes of extracted features span from the most well known such as statistical/spectral kurtosis, mean and median of a signal, to less frequently employed features such as human range energy ratio, the estimator of the cumulative distribution function (ECDF), variability and peak-to-peak. The idea behind considering a large number of initial features was to assess their individual importance, and subsequently retain only the most informative ones by adopting a systematic feature selection approach rather than selecting them subjectively from intuition.
Feature reduction can also facilitate model training in terms of required computational time. The selection criteria were based on the ``SelectKBest'' feature selection algorithm that ranks the features with the largest ANOVA F-value between a feature vector and a class label. The reason for choosing such an approach is that it offers a better trade-off in terms of accuracy, stability and stopping criteria in comparison to other feature selection algorithms such as SelectPercentile or VarianceThreshold \citep{powell2019cross}. Determining the threshold for an optimal number of features is an iterative process of supervised evaluation of model performance with variable numbers of features. Initially, a model with all features was built and its performance metric in terms of accuracy was observed and in subsequent steps, the number of features was reduced by half iteratively as long as the model performance increased. The following iterative step with number of features that would indicate a decrease in model performance served as a stopping criteria. Finally, a Pearson correlation was computed to assess the correlation between selected features in order to reduce multicollinearity among them. Reducing multicollinearity of features is an essential step for retaining the predictive power of each of them. Using  highly correlated features very often hamper model training. Experiments conducted by \cite{lieberman2014precise} indicate a correlation threshold of $\pm0.5$ for optimal model performance.

\subsection{Models Training}
The modelling and training process aims at learning classification models capable of discriminating self-reported mental workload scores from subjects (target feature), given the features extracted and selected in the previous step D (independent features). The mental workload scores were selected rather than the type of condition (``Simkap'' or ``Rest'') because we wanted a sensible indicator of mental workload, not a task load condition. In other words, a self-reported indicator of mental workload can be considered a more reliable representation of the user experience than a class representing a certain task load condition. This argument is originated by the fact that, in both task load conditions, users can experience any level of cognitive load. For example, a novice user can experience high mental workload for an easy task load condition when compared to an expert user. Similarly, a skilled user can experience moderate mental workload even while in a resting condition because of significant mind wandering. In research from \cite{charles2019measuring}, a distinction between objective elements of the work (taskload) from the subjective perception of mental workload is outlined. Both taskload and subjective perception of mental workload can be mediated by operator experience or time constraint factors. 
Therefore, it is intuitive that task load conditions are not equivalent to mental workload experiences. In fact, on one hand, the former are strictly defined prior task execution, and are static, meaning they are immutable during task execution. 
On the other hand,  the latter are unknown prior task execution and can change depending on a number of factors, for example including user's prior knowledge, motivation, time of execution, fatigue, stress among the others. To stress further, research has clearly shown that even the same person can execute a task, designed with a specific, static load condition (pre-defined task demands) differently at various times of the day \citep{hancock1992effect}.

Additionally, to facilitate subsequent interpretation, we treated model training as a binary classification problem, mainly to use more interpretable evaluation metrics such as precision, recall, accuracy and f1 score. Therefore, the target feature range of 1 to 9 of the self-reported mental workload scores was mapped into two levels of mental workload, the ``suboptimal MWL'' and ``super optimal MWL''. The split was adopted based on the assumption of the parabolic relationship between experienced mental workload and performance as outlined in \cite{Longo2021}. This split was done by aggregating the scores from 1 to 4, representing some degree of low mental workload (effort), into the ``suboptimal MWL'', and all the scores from 6 to 9, for all of those supporting some degree of high mental workload (effort), into the ``super optimal MWL''. All those scores rated five were discarded because they represent the neutral experience of mental workload.

The learning techniques chosen for achieving this aim are Logistic regression (L-R), Support Vector Machines (SVM) and Decision Trees (DTR). Many research works have considered these three learning techniques for continuous and more prolonged EEG recordings \citep{berka2007eeg, doma2020comparative, hu2018automated}. Logistic regression and SVM, as error-based learning techniques, are suitable for binary classification tasks (as in this work). On the other hand, as an information-based technique, decision trees are suitable for distinguishing important features by calculating their information gains during model training. 

Due to the fact that a small dataset of 48 subjects was selected, then repeated montecarlo sampling for model training and validation is set in the following order:
\begin{enumerate}
    \item a randomised 70\% of subjects is selected both from the ``suboptimal MWL'' and the `super optimal MWL'' classes (dependent feature) for model training;
    \item the remaining 30\% was kept for model testing.
    \item the above splits are repeated for 100 iterations to observe random training data, and effectively capture the probability density of the target variable.
\end{enumerate}
A general rule of thumb implies a minimum of $1/5th$ ratio for each feature in the data to increase model accuracy \citep{friedman1997bias}. Given the low number of training instances in each of the target classes (``suboptimal MWL'', ``super optimal MWL''), the ``curse of the dimensionality'' problem is anticipated \citep{verleysen2005curse}. 
Therefore, a strategy for generating synthetic data is adopted, which is based on the generation of statistically similar synthetic data that mimics the original data. For this purpose, the Synthetic Data Quality Score based on metrics like Field Correlation Stability, Deep Structure Stability and Field Distribution Stability \citep{GretelAI2021} is adopted.

The Field Correlation Stability is the correlation between every pair of independent features (fields) in the training data and then in the synthetic data. These values' absolute difference is computed and averaged across all independent features. The lower this average, the higher is the correlation stability of the synthetic data.
Deep Structure Stability verifies the statistical integrity of the generated dataset by performing deep, multi-field analysis of distributions and correlations. This is done by executing Principal Component Analysis (PCA) on the original data, and comparing it against that on the synthetic data. A synthetic quality score is created by comparing the distributional distance between the principal components found in the two datasets. The closer the principal components, the higher the quality of the synthetic data.
Field distribution stability measures how closely the field distribution in the synthetic data mimic that on the original data. The comparison of two distributions is done using the Jensen-Shannon (JS) distribution distance given as:
    \begin{equation}
    JSD = H(M) - \frac{1}{2}(H(O) - H(S))
    \end{equation}
where $H(O)$ and $H(S)$ are the Shannon entropy values for original (O) and synthetic (S) data respectively and $H(M)$ is the sum of selected weights for probability distributions ($\pi$) and dataset probabilities ($P$) given as $M = \sum_{i=1}^{2}\pi_iP_i$. The lower the distance score on average across all fields, the higher the Field Distribution Stability quality score and consequently the higher the quality of the synthetic data generated.

The Synthetic Data Quality Score represents an arithmetic mean between field correlation stability, deep structure stability and field distribution stability. In this sense, the Synthetic Data Quality Score can be viewed as a confidence score as to whether scientific conclusions drawn from the synthetic dataset would be indistinguishable as if they were to be used in the original data. Synthesizing new data is performed using synthetic generators offered from Gretel.ai\footnote{Gretel.ai - Privacy Engineering as a Service for Data Scientists - https://gretel.ai}. The training process for the combined (original + synthetic) uses the same montecarlo sampling with the same steps as with original data outlined above. Randomised 70\% of subjects is selected both from the combined (original + synthetic) for the``suboptimal MWL'' and the `super optimal MWL'' classes (dependent feature) for model training, the remaining 30\% of combined (original + synthetic) subjects was kept for model testing and performing 100 iteration through these randomized splits. During model training, the data was normalized using $\varB{z-score}$ normalization given as $z=(x-\mu)/\sigma$, where $\mu$ is the mean of training samples and $\sigma$ is the standard deviation. The rationale for using $\varB{z-score}$ normalization is that it tends to minimize the mean ($\mu$=0) and maximize the standard deviation ($\sigma=1$) for the normalized value and makes it suitable reducing extremely peak values in data, by transforming it in such a way that it’s no longer a massive outlier.  

\subsection{Models Evaluation}
A set of evaluation metrics were employed to assess the ability of the selected models to generalize on unseen data by learning from the training data. These metrics can be used to measure and summarise the quality of the trained models when tested with previously unseen data. For a binary classification problem, such as in the case, the evaluation of the models is dependent on \textit{True Positives} $(tp)$ and \textit{True Negatives} $(tn)$ which denote the number of positive and negative instances that are correctly classified. It can be also conducted with the \textit{False Positives} $(fp)$ and \textit{False Negatives} $(fn)$ that denote the number of miss-classified negative and positive instances, respectively. According to this, several metrics are used to evaluate the performance of the trained models. \textit{The accuracy metric} measures the ratio of correct predictions over the total number of evaluated instances. Accuracy is represented as, $Accuracy = (tp+tn)/(tp+fp+tn+fn)$.\textit{Precision} is used to measure the positive instances that are correctly predicted from the total predicted instances in a positive class, given as $Precision = (tp)/(tp+fp)$. \textit{Recall} measures the fraction of positive instances that are correctly classified, $Recall = (tp)(tp+tn)$. \textit{F-Measure} or \textit{f1-score} is the harmonic mean between recall and precision values represented as, $\varB{f1-score} =(2\cdot Precision \cdot Recall)/(Precision+Recall)$.
The proposed evaluation metrics are essential to assess the robustness of the selected models built upon high-level features extracted from the MWL indexes towards the discrimination of self-reported perceptions of mental workload. While precision refers to the percentage of relevant instances, recall refers to the rate of total relevant instances correctly classified by the model. The best model minimizes the value of $(fp)$ in precision and $(tn)$ in recall, and both come at the cost of each other since we cannot minimize both of them in one metrics. f1-score represent a harmonic mean of precision and recall and takes into account both metrics. Consequently, to bring hypotheses H1 and H2 on provable grounds, the $\varB{f1-score}$ metric is adopted too.

\renewcommand\thesubsection{\Roman{subsection}}
\section{Results}
\label{sec:results}
The results section follows the same order of steps as outlined in the design section. 

\subsection{EEG artefact removal} 
\begin{figure}[!ht]
\centering
\includegraphics[width=13cm]{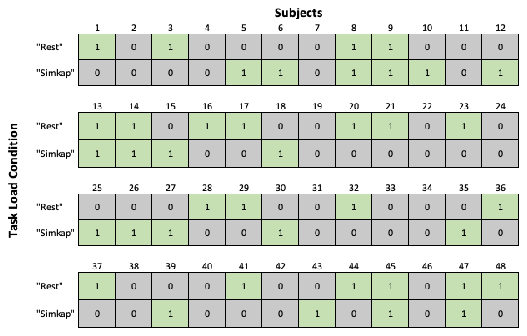}
\caption{The number of components removed across all 48 subjects for ``Rest'' and ``Simkap'' task load conditions.}
\label{fig:0}
\end{figure}

Artefact removal is performed on each EEG signal for each of the 48 subjects separately for the ``Rest'' and ``Simkap'' task load conditions. The average number of ICA components removed from the EEG data associated with each subject is $1.61$ for the ``Rest'' and $1.46$ for the ``Simkap'' condition. The number of removed artefacts is within limits of the adopted methodology described in \citep{NOLAN2010152}. Figure \ref{fig:0} depicts the removal occurrence for a total of 14 components across all 48 users for ``Rest'' condition and ``Simkap'' condition. As it is possible to see from figure \ref{fig:0}, at most, one ICA component per subject that is significantly different from the other components ($\pm3$ standard deviations) exists. These components are removed by zero-ing them, and the EEG multi-channel data is subsequently reconstructed by applying inverse ICA. Since at least one bad component was identified and removed for most subjects, it is possible to reasonably claim that some artefact has been removed from the EEG signal, thus facilitating the subsequent computations of the alpha and theta bands.

\subsection{Evaluation of Feature Extraction and Selection}
\begin{figure}[!ht]
    \includegraphics[width=9cm]{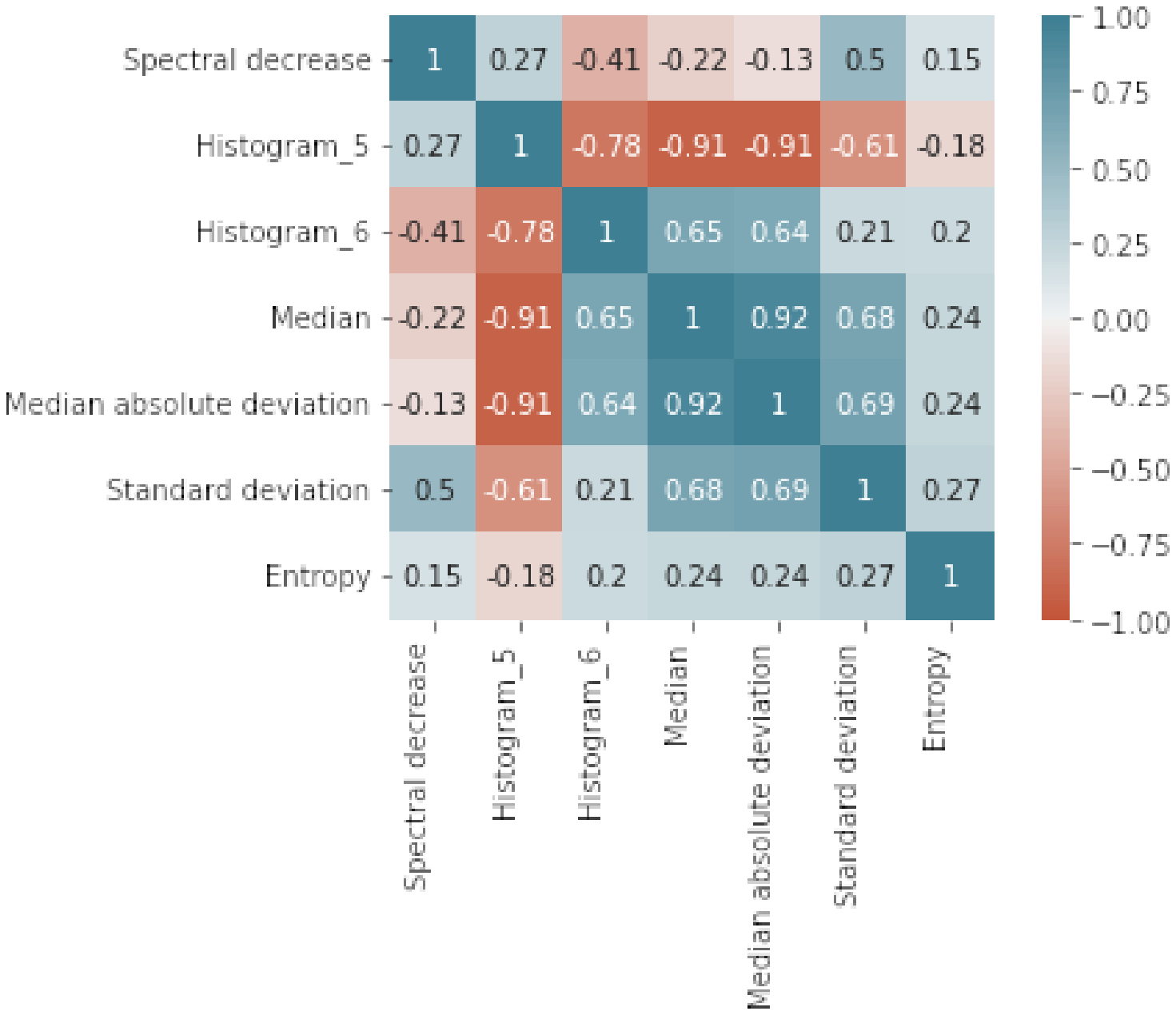}
    \includegraphics[width=9cm]{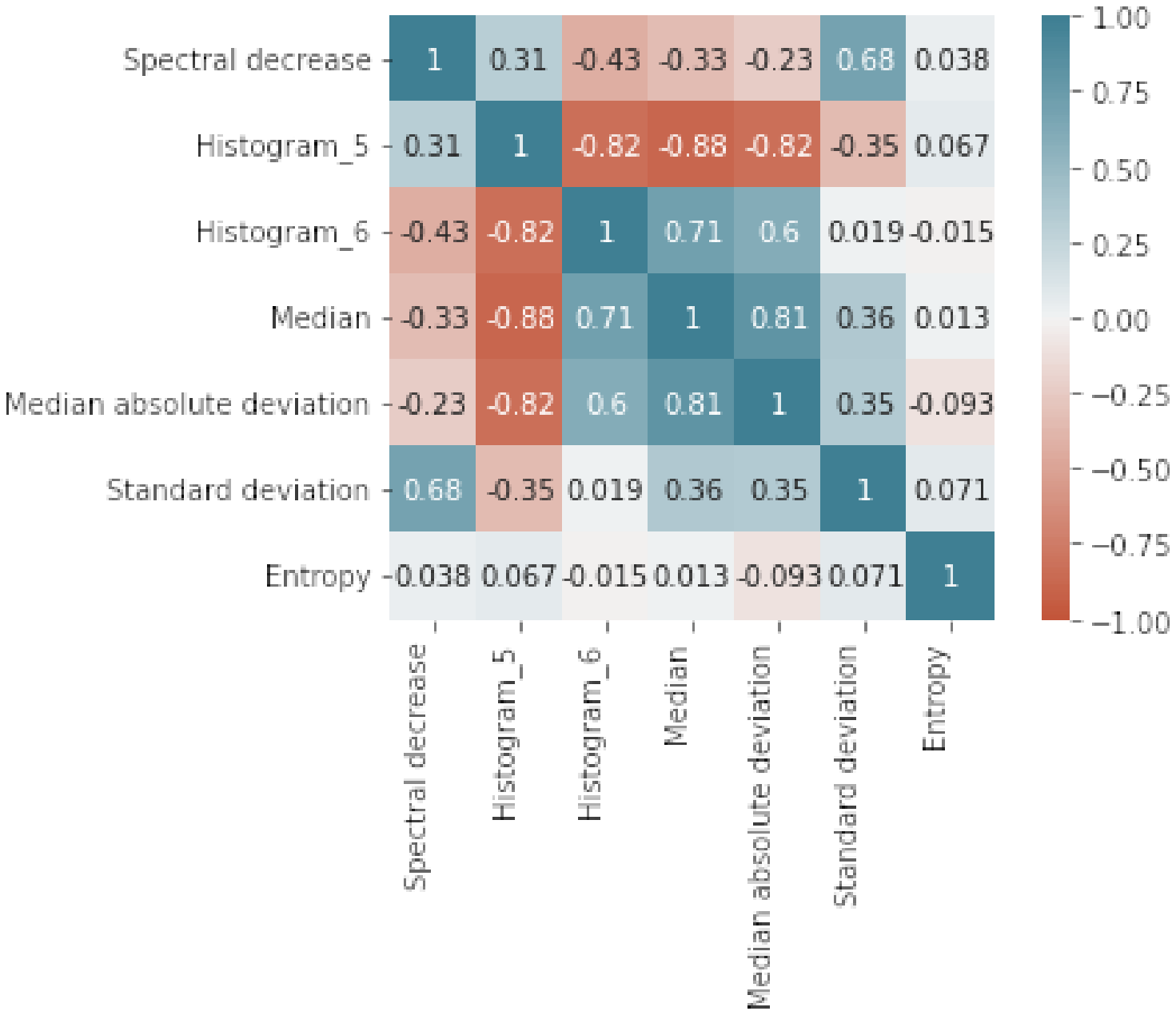}
    \caption{Pearson correlation coefficients matrix for the case of MWL index - \textbf{at-1} - Rest (left) and Simkap (Right) task-load conditions. \textbf{at-1}: alpha-to-theta ratios between the indexes $\varB{c-\alpha}$ and $\varB{c1-\theta}$. The scale on the right of the image indicates the Pearson correlation coefficients range.}
    \label{fig:3}
\end{figure}

All high-level features are collected from the statistical properties of the mental workload indexes in various domains, including the temporal and frequency domains. The initial number of collected features are 210, and the exhaustive list is provided in the supplementary material accompanying this article. The ANOVA F-Value is computed for each of these features, and those with the highest value are retained for subsequent model training.  
\begin{figure}[!ht]
    \centering
    \includegraphics[width=13cm]{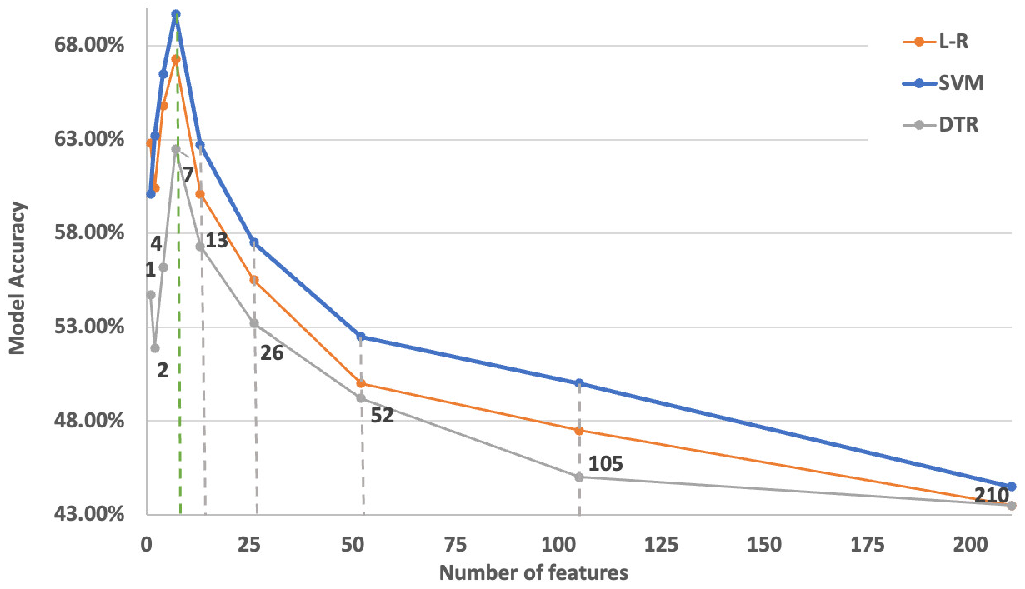}
    \caption{Optimal number of features against model performance with data coming from the Simultaneous Task EEG workload (STEW) dataset. The dashed lines indicate the number of feature considered in the iteration. It can be seen that the optimal number of top features to select is around seven indicated with green dashed line which also acts as a stopping criteria.}
    \label{fig:3_1}
\end{figure}
Since the SelectKBest algorithm requires an initial number of features, an iterative approach of feature inclusion during model training and the performance of the models with those features is assessed through its accuracy. Iterative optimal feature selection is performed by employing data from the original dataset.
Figure \ref{fig:3_1} illustrates the convergence on the optimal number of features in relation to model performance grouped by learning techniques (L-R, SVM, DTR). This resulted in a reduced number of features that are kept for model training by employing to the dataset enhanced with synthetic data. Figure \ref{fig:3} shows the Pearson correlation matrix for the ``Rest'' and the ``Simkap'' states for the alpha-to-theta ratios for the case of index $at$-$1$ (as designed in section \ref{sec:design}C).
Noticeably, most of the features are in the correlation range between $-0.5$ and $+0.5$, which contributes to reduce multicollinearity and thus potentially being all relevant and with high prediction capability \citep{lieberman2014precise}. Figure \ref{fig:3} is an illustration of the results associated to a single mental workload  index (at-1). However, results associated to the other indexes are mostly consistent with these, as it is possible to examine in the supplementary figures (S.1 to S.9) accompanying this article.

\subsection{Evaluation of the training set across indexes}
After the feature selection process, training of the models was conducted with Montecarlo sampling using Logistic Regression (L-R), Support Vector Machines (SVM) and Decision Trees(DTR) as described in design subsection E). Model training suffered from the ``curse of dimensionality'' issue since it comprised 48 subjects across only the seven selected features. The number of training instances is low compared to the number of independent features retained for modelling purposes. This is followed by the peak phenomenon of feature inclusion, where the number of features and their cumulative discriminatory effect is essential for the average predictive power of a classifier, which is data-dependent \citep{zollanvari2019theoretical}. 
\begin{figure}[!ht]
\centering
\captionsetup{justification=centering}
\includegraphics[width=8.8cm]{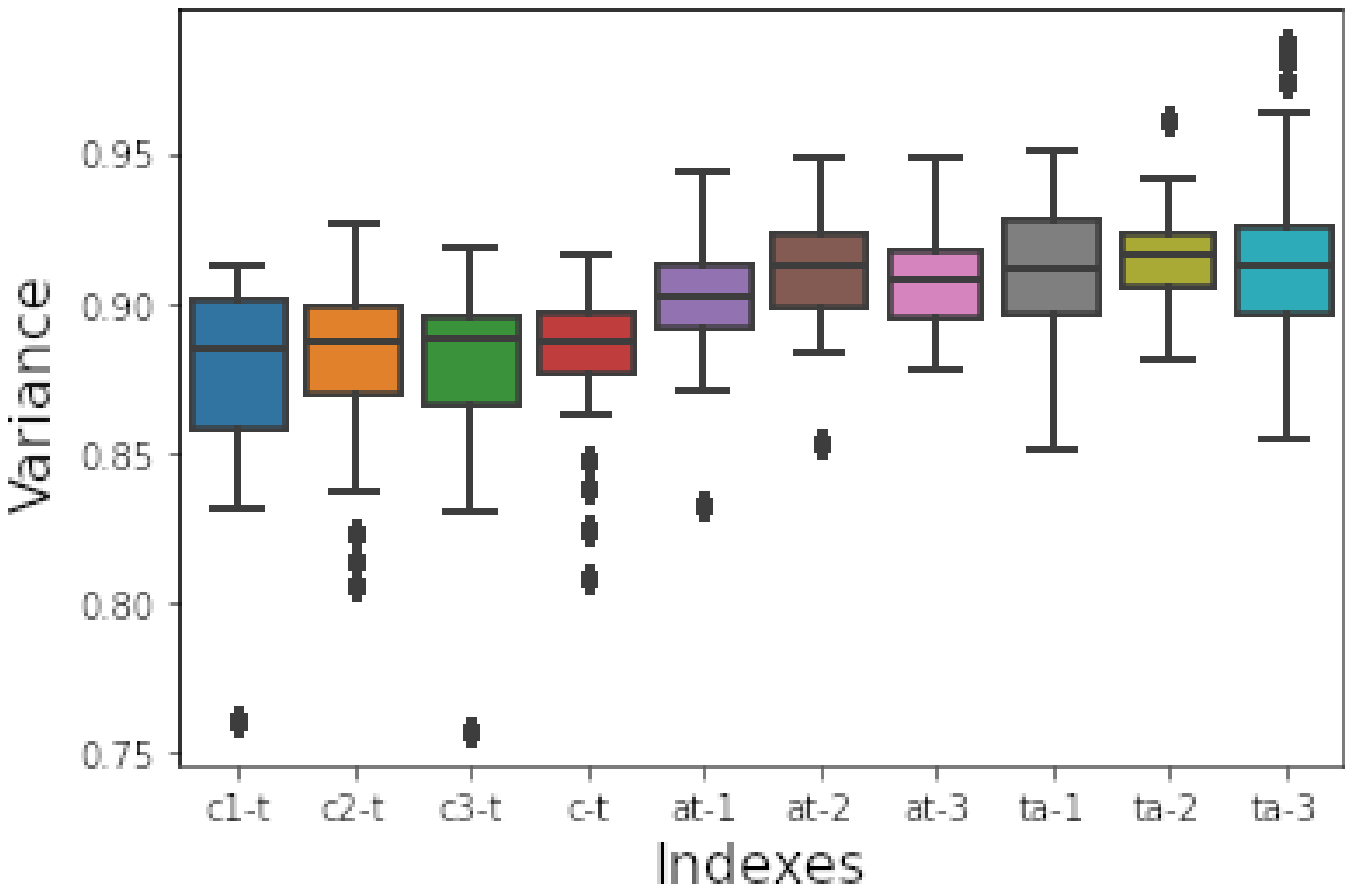}
\includegraphics[width=8.8cm]{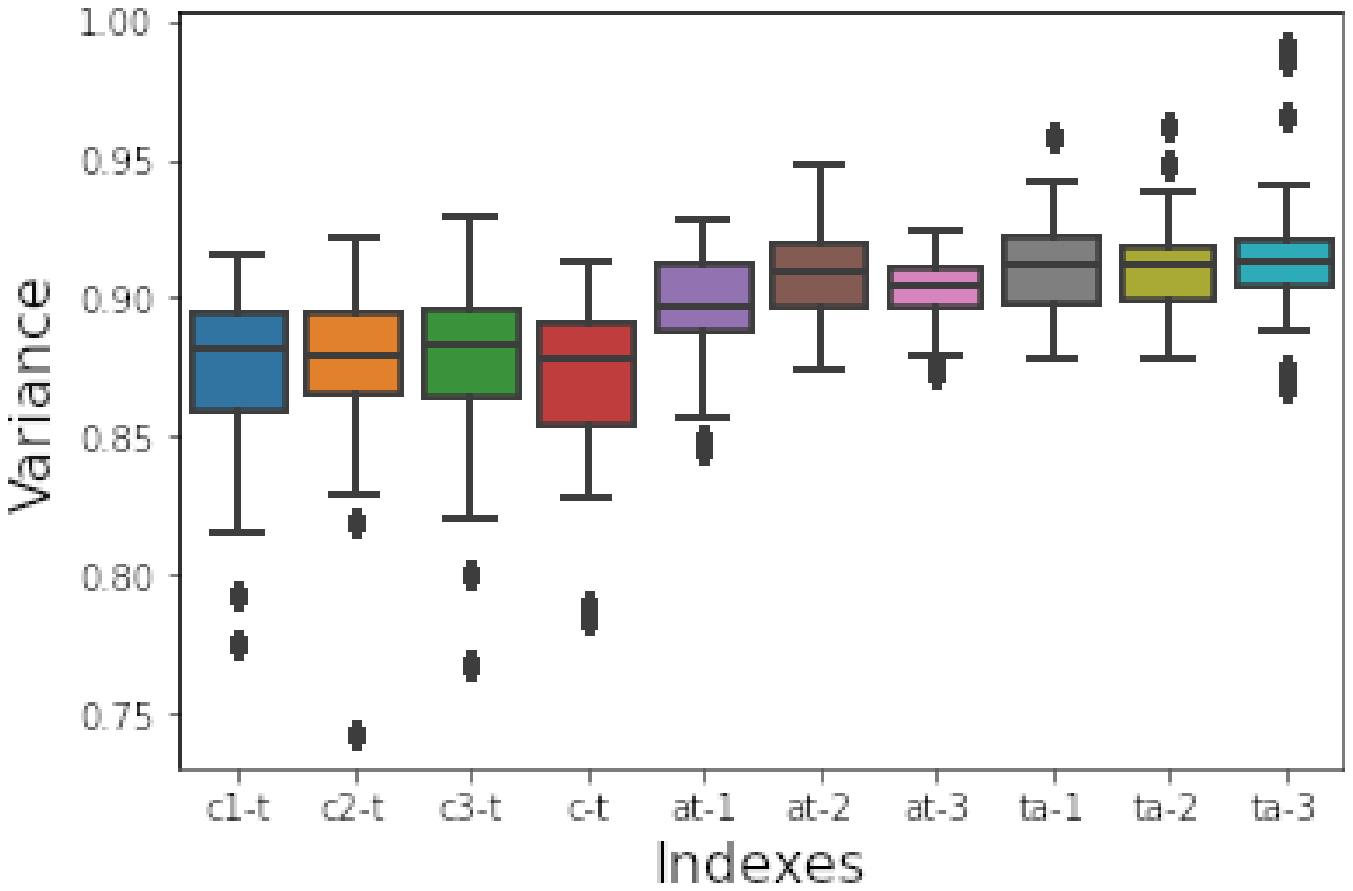}
\caption{Variance of spectral entropy associated the original data - Left (``Rest'' state), Right (``Simkap'' state). From the figure it can be seen a small interquartile range Q1- Q3 is small.}
\label{fig:2}
\end{figure}
The initial model evaluation with test data on the original dataset did not reveal the accuracy of more than 60\% for the standalone mental workload indexes built upon the alpha and theta indexes alone ($\varB{c1-\theta}$, $\varB{c2-\theta}$, $\varB{c3-\theta}$, $\varB{c-\alpha}$).
An accuracy of 70\% for the mental workload indexes built upon the alpha-to-theta and theta-to-alpha ratios was observed. 
An in-depth analysis of the learning curves  associated to the classifiers indicated a model underfitting and an inability to generalize from test data. Moreover, analyzing the spectral entropy of the mental workload indexes revealed a small variance variation, as can be seen from the boxplots of figure \ref{fig:2}. Small data variance subsequently increases the bias influencing the model's ability to generalize. Thus, as expected, synthetic data generation was applied for training robust models.

\subsection{Synthetic data evaluation}
The input for data synthesis was the initial dataset comprised of 48 subjects and 150 data points (2.5 minutes of EEG data per participant split into segments of 1 second) for each of the indexes designed in equation \ref{eq:eq-3}. Two synthetic datasets are created, one for the ``Rest'' and one for the ``Simkap'' task load conditions in order to retain the original dataset's intrinsic properties. Table \ref{tab:table-2} illustrates the overall synthetic quality scores for the mental workload indexes set in equation \ref{eq:eq-3}.
\begin{table}[!ht]
\centering
\caption{\label{tab:table-2}Synthetic score for different mental workload indexes and two task load conditions (``Rest'' and ``Simkap''}
\vspace*{1mm}
\renewcommand{\arraystretch}{1.1}%
\begin{tabular}{p{2cm} p{1cm} p{1cm} p{1cm} p{1cm} p{1cm} p{1cm} p{1cm} p{1cm}} 
\hline
 MWL & \multicolumn{2}{c}{Field Corr. Stability} & \multicolumn{2}{c}{Deep Struct. Stability} & \multicolumn{2}{c}{Field Dist. Stability} & \multicolumn{2}{c}{Overall Score}\\ [0.5ex] 
 index: & Rest & Simkap & Rest & Simkap & Rest & Simkap & Rest & Simkap \\
 \hline\hline
 $\varB{c1-\theta}$ & 100\% & 100\% & 95\% & 98\% & 78\% & 70\% & 91\% & 89\% \\
 $\varB{c2-\theta}$ & 100\% & 96\% & 94\% & 94\% & 78\% & 71\% & 91\% & 87\%\\
 $\varB{c3-\theta}$ & 94\% & 95\% & 97\% & 100\% & 78\% & 80\% & 89\% & 91\% \\
 c-$\alpha$ & 100\% & 94\% & 97\% & 100\% & 72\% & 81\% & 90\% &	91\% \\
 $\varB{at-1}$ & 91\%	& 94\%	& 89\%	& 92\%	& 73\% & 74\% &	84\% & 87\% \\
 $\varB{at-2}$ & 100\% & 92\% & 93\% & 94\% &76\% & 78\% & 90\% & 88\% \\
 $\varB{at-3}$ & 93\% & 100\% & 95\% & 94\% & 72\% & 74\% & 87\% & 89\% \\
 $\varB{ta-1}$ & 100\% & 100\%	& 93\% & 95\% &	77\% &	75\% &	90\% &	90\% \\
 $\varB{ta-2}$ & 91\% & 100\% &	96\% & 88\% & 76\% & 73\% & 88\% & 87\% \\
 $\varB{ta-3}$ & 100\% &	96\% &	85\% &	94\% &	80\% &	82\% &	88\% &	90\% \\
 \hline
\end{tabular}
\end{table}
Findings suggest that the overall synthetic data score is always above 87\% throughout all the mental workload indexes selected for the comparative analysis. The synthetic quality score was measured in the scale (1-20)\%-Very Poor, (20-40)\%-Poor, (40-60)\%-Moderate, (60-80)\%-Good and (80-100)\%-Excellent. This suggests that the quality of synthesised data is excellent, and in line with similar studies \citep{hernandez2020novel}. Consequently, the synthesised data was of an additional 180 synthesised subjects with 150 data points (2.5 minutes of EEG activity split into 150 segments of 1 second) for each mental workload index. The final combined dataset with original and synthesized data is now comprised of 228 subjects with 150 data points for each mental workload index, as defined in set \ref{eq:eq-3}.

\subsection{Validation of models for discriminating self-reported perceptions of mental workload}

The training of the models with Logistic Regression and Support Vector Machines learning techniques utilized the linear optimizer since it offers speed and optimum convergence on minimizing a multivariate function by solving univariate optimization problems during repeated training of the model \citep{fan2008liblinear}. In the case of model training with Decision Tree, a Gini index was used to measure the quality of split during the model build. 

\begin{figure}[!ht]
    \includegraphics[width=9cm]{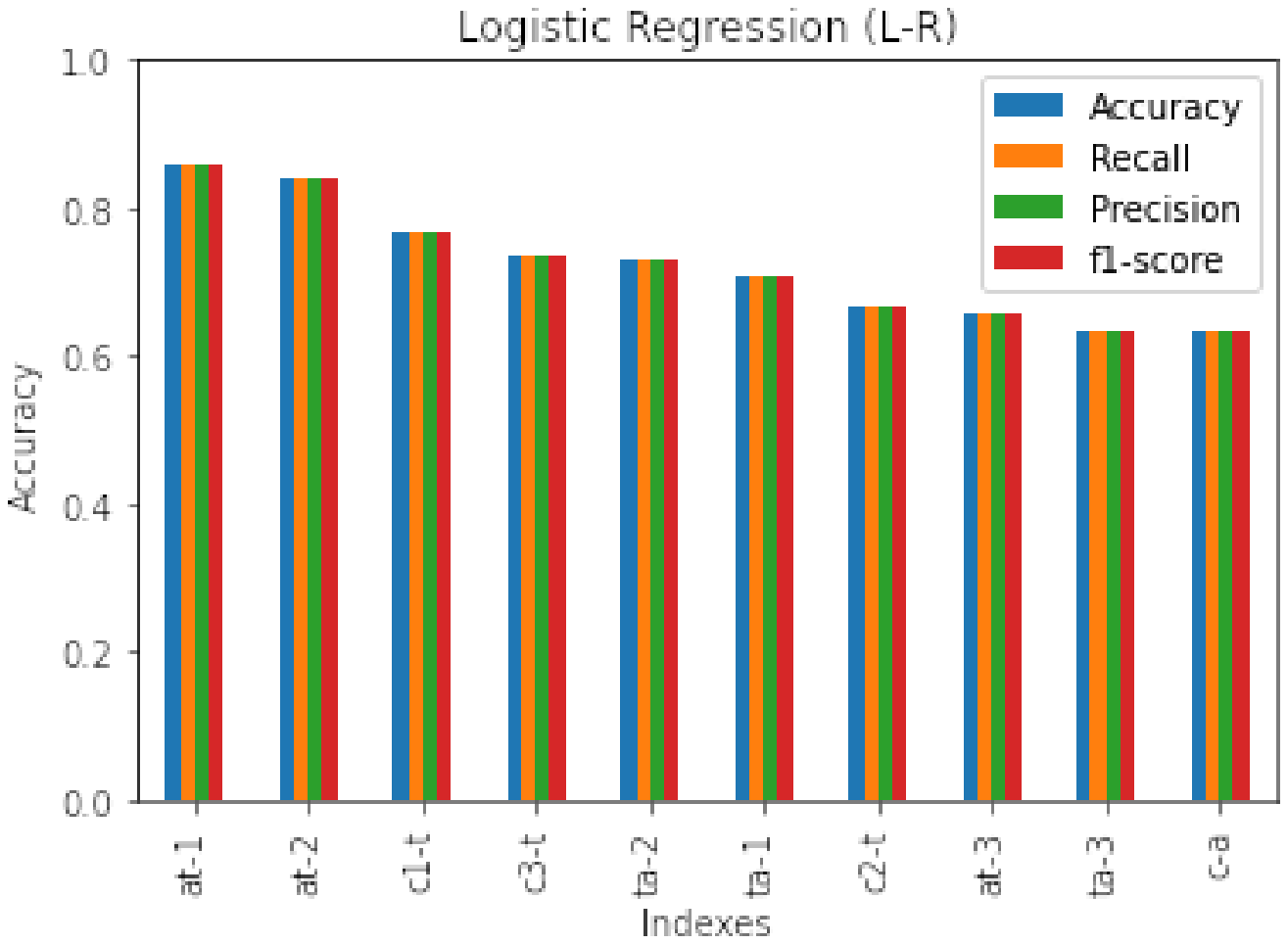}
    \includegraphics[width=9cm]{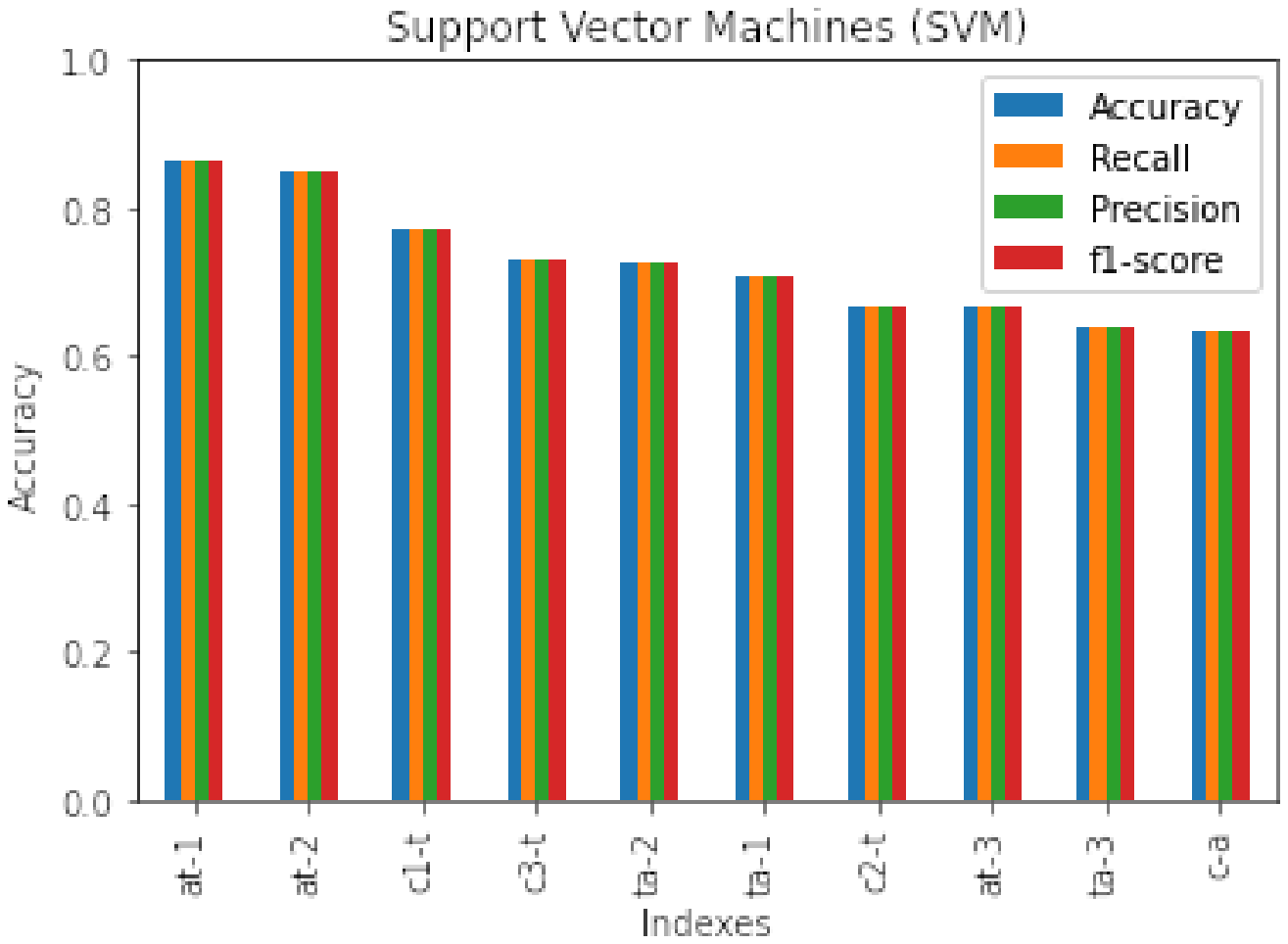}
    \hfill
    \includegraphics[width=9cm]{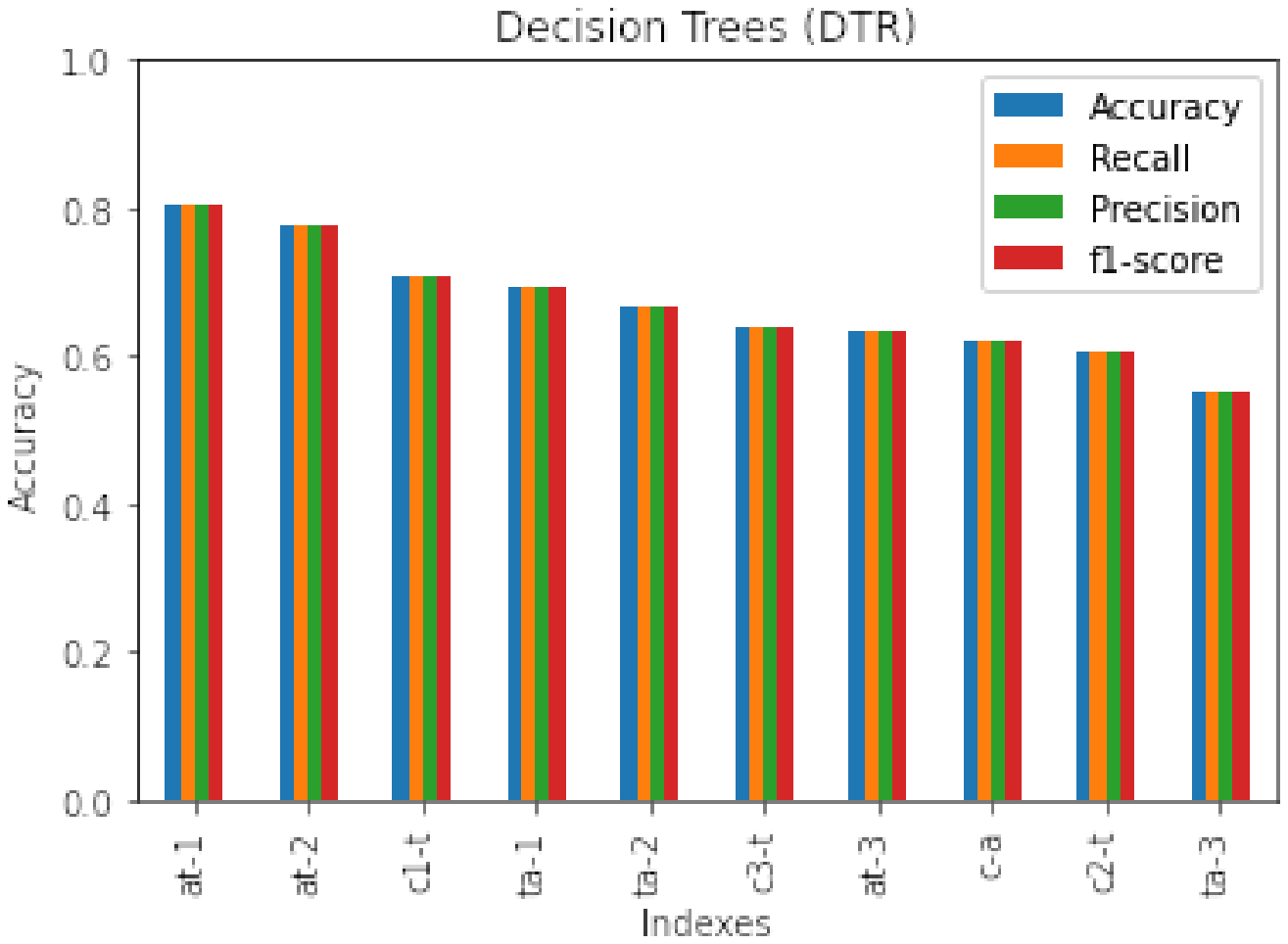}
    \caption{Classification results with high-level features across models with different learning techniques.}
    \label{fig:4}
\end{figure}

\begin{figure}[!ht]
    \includegraphics[width=18cm]{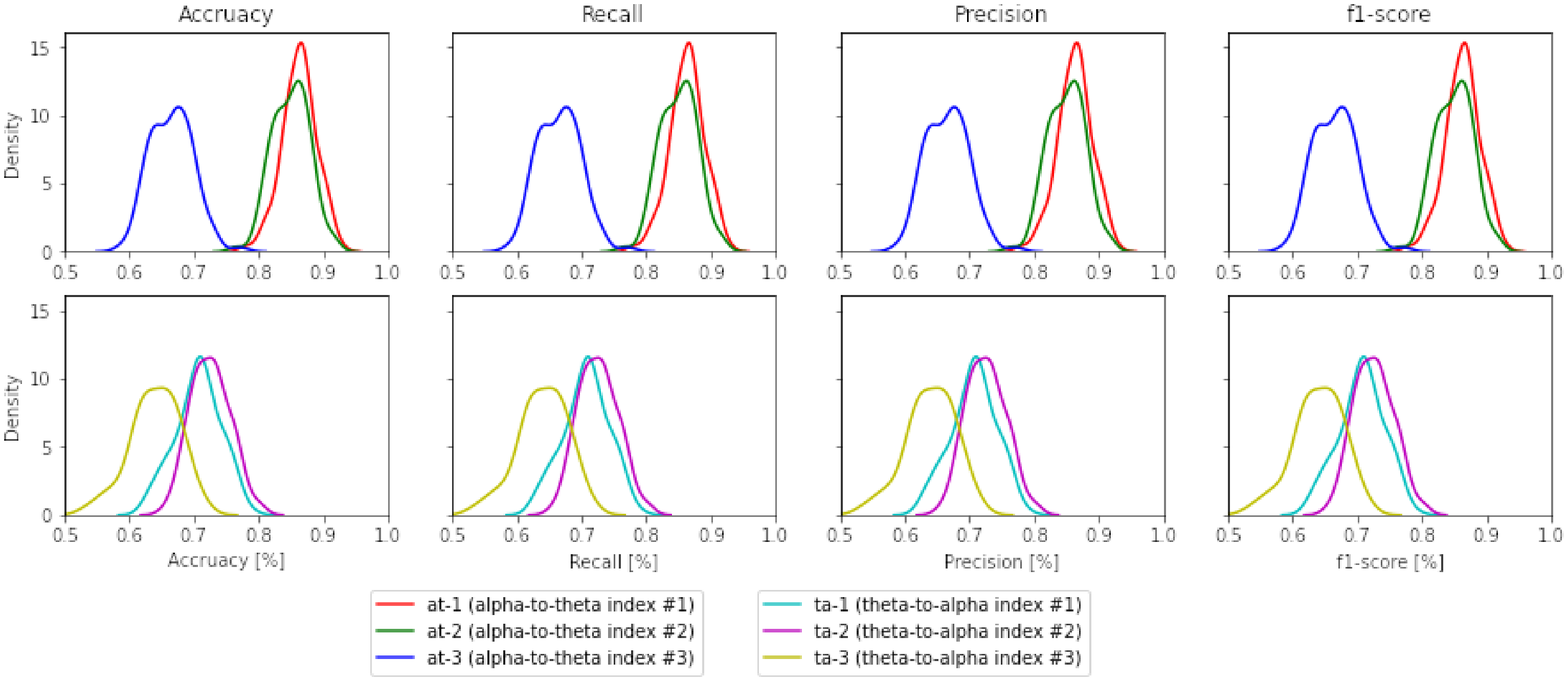}
    \caption{Density plots across all performance metrics between all band ratio indexes (the case for SVM)}
    \label{fig:6}
\end{figure}

The classifiers performance is shown in figure \ref{fig:4}. The evaluation metrics are shown across all mental workload indexes and are presented in descending order. 
The best classification accuracy results are observed for those models built with Support Vector Machines (SVM) and Logistic regression (L-R). In order to acknowledge the best learning technique, a two-tailed t-test between the three learning techniques and the employed evaluation metrics was performed. The results indicated no statistically significant difference between Logistic Regression (L-R), Support Vector Machines (SVM) or Decision Trees (DTR). This indicates the validity of the training approach adopted from the design, which means that no matter the learning technique adopted, the results across all applied evaluation metrics are the same. Table \ref{tab:table-3} illustrates the p-value significance levels of the t-test between evaluation metrics for each learning technique used in the study. The t-test was conducted with a threshold confidence value of $\alpha=0.05$.
\begin{table}[!ht]
\centering
\caption{\label{tab:table-3} The two-tailed t-test between L-R, SVM and DTR and f1-score, accuracy, recall and precision}
\vspace*{1mm}
\renewcommand{\arraystretch}{1.1}%
\begin{tabular}{c c c c} 
\hline
 & \textit{L-R - SVM} & \textit{SVM-DTR} & \textit{L-R - DTR} \\[0.5ex]
\hline\hline
& \multicolumn{3}{c}{t-statistic (p-value)}\\
\hline
f1-score & -0.04 (0.96) & 1.55 (0.13) & 1.50 (0.15) \\
accuracy & -0.05 (0.96) & 1.55 (0.14) & 1.50 (0.15) \\
precision & -0.04 (0.96) & 1.55 (0.14) & 1.50 (0.14) \\ 
recall & -0.04 (0.96) & 1.55 (0.14) & 1.51 (0.15) \\
 \hline
\end{tabular}
\end{table}
Further analysis of mental workload indexes of the alpha-to-theta ratios ($\varB{at-1}$, $\varB{at-2}$) indicates better performance than their respective individual counterparts used for computing those ratios ($\varB{c1-\theta}$, $\varB{c2-\theta}$ and $\varB{c-\alpha}$). For example, in the case of all learning techniques (L-R, SVM and DTR), first two alpha-to theta ratio indexes ($\varB{at-1}$ and $\varB{at-2}$) show better performance than their individual counterparts ($\varB{c1-\theta}$, $\varB{c2-\theta}$ and $\varB{c-\alpha}$). 
\begin{figure}[!ht]
    \includegraphics[width=18cm]{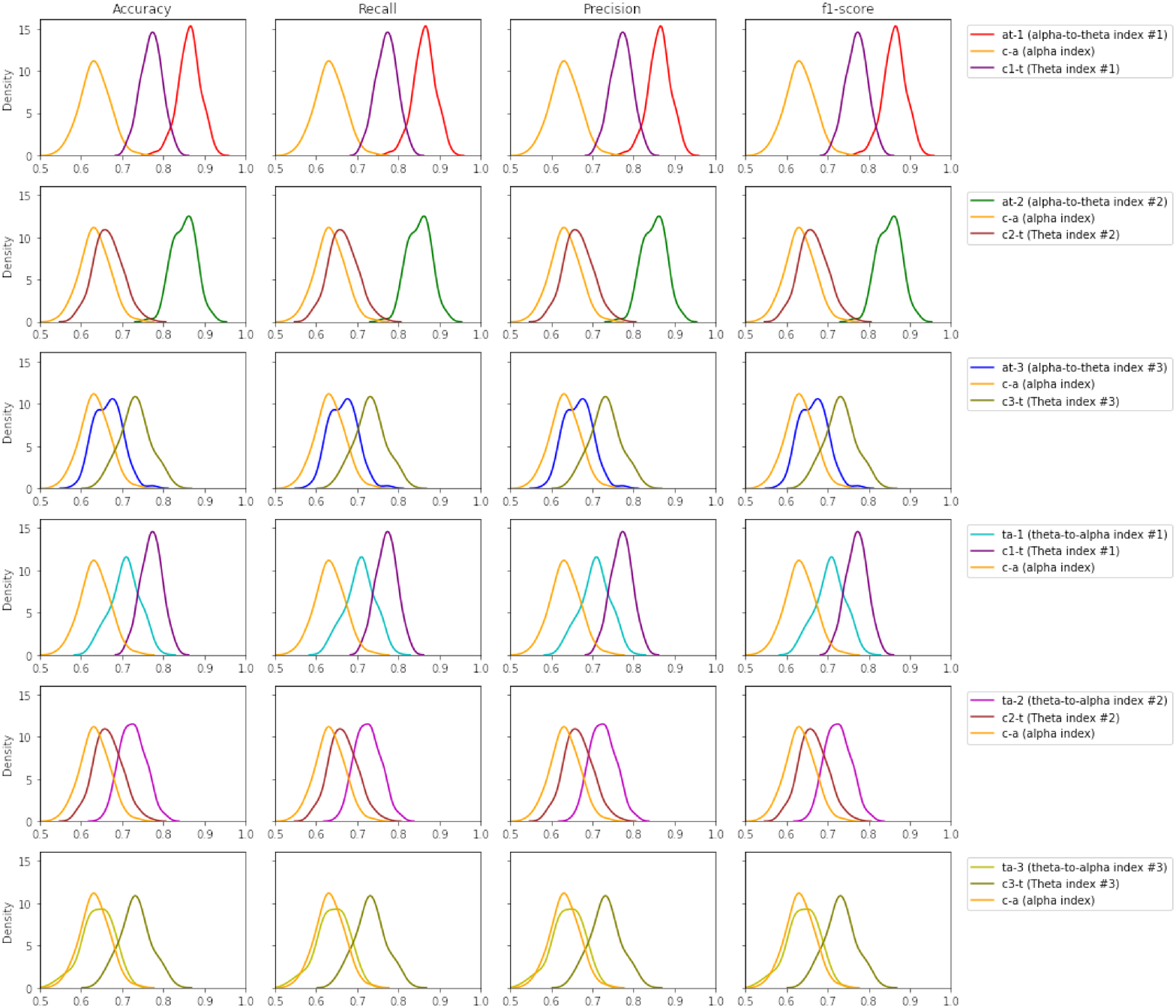}
    \caption{Density plots for SVM across all performance metrics between band ratio indexes vs. their individual indexes}
    \label{fig:5}
\end{figure}
In the case of the theta-to-alpha mental workload indexes, this is also seen in the first two indexes ($\varB{ta-1}$ and $\varB{ta-2}$). Figure \ref{fig:5} illustrates the performance of band ratios (alpha-to-theta and theta-to-alpha) across all evaluation metrics, given as density plots for the case of Support Vector Machines (SVM). The density plots for all other learning techniques are available on the supplementary materials (figures S.10 and S.11). Table \ref{tab:table-4} outlines the significance levels of a two-tailed t-test between the alpha-to-theta and theta-to-alpha ratio indexes with indexes used to construct those ratios.
\begin{table}[!ht]
\centering
\caption{\label{tab:table-4} The two-tailed t-test between the alpha-to-theta and theta-to-alpha ratio indexes with their individual indexes with Bonferroni(\dag) correction applied, resulting in a significance level set at $\alpha=0.005$}
\vspace*{1mm}
\renewcommand{\arraystretch}{1}%
\begin{tabular}{c c m{3.5cm} m{3.5cm} m{3.5cm} m{3.5cm}} 
\hline
& & $\varB{c1-\theta}$ & $\varB{c2-\theta}$ & $\varB{c3-\theta}$ & $\varB{c-\alpha}$ \\ [0.5ex] 
\hline\hline
& & \multicolumn{4}{c}{t-statistic (p-value)}\\
\hline
\multirow{ 6}{*}{L-R} & $\varB{at-1}$ & 25.1 $(1.96 \cdotp 10^{-63})^{**}$ $(2.36 \cdotp 10^{-62})^{\dagger}$	& - & -	& 51.4 $(1.32 \cdotp 10^{-111})^{**}$ $ (1.56\cdotp 10^{-115})^{\dagger}$\\
& at-2 & - & 42.3 $(3.94\cdotp 10^{-101})^{**}$ $(4.73\cdotp 10^{-100})^{\dagger}$ & - & 46.8 $(4.82\cdotp 10^{-109})^{**}$ $(5.78\cdotp 10^{-108})^{\dagger}$\\
& at-3 & -	& - & 15.5 $(3.93\cdotp10^{-36})^{**}$ $(4.71\cdotp 10^{-35})^{\dagger}$ & 4.90 $(1.95\cdotp10^{-6})^{**}$ $(2.34\cdotp 10^{-5})^{\dagger}$\\
& ta-1 & 14.3 $(1.49\cdotp10^{-32})^{**}$ $(1.79\cdotp 10^{-31})^{\dagger}$	& - & -	& 15.1 $(4.40\cdotp 10^{-35})^{**}$ $(5.28\cdotp 10^{-34})^{\dagger}$\\
& ta-2 & - & 14.9 $(2.96\cdotp 10^{-34})^{**}$ $(3.56\cdotp 10^{-33})^{\dagger}$ & -	& 20.7 $(1.07\cdotp 10^{-51})^{**}$ $(1.28\cdotp 10^{-50})^{\dagger}$\\
& ta-3 & -	& - & 19.4 $(5.81\cdotp 10^{-48}$) $(6.97\cdotp 10^{-47})^{\dagger}$	& 0.49 (0.62) $(1)^{\dagger}$\\
\hline
\multirow{ 6}{*}{SVM} & at-1 & 25.0 $(1.99\cdotp 10^{-63})^{**}$ $(2.39\cdotp 10^{-62})^{\dagger}$	& - & -	& 52.4 $(4.40\cdotp 10^{-118})^{**}$ $(5.28\cdotp 10^{-117})^{\dagger}$\\
& at-2 & - & 42.2 $(4.48\cdotp 10^{-99})^{**}$ $(5.37\cdotp 10^{-98})^{\dagger}$ & - & 48.1 $(2.69\cdotp 10^{-111})^{**}$ $(3.22\cdotp 10^{-110})^{\dagger}$\\
& at-3 & -	& - & 13.2 ($4.75\cdotp10^{-29})^{**}$ $(2.79\cdotp 10^{-28})^{\dagger}$ & 6.68 ($2.39\cdotp 10^{-29})^{**}$ $(5.71\cdotp 10^{-28})^{\dagger}$\\
& ta-1 & 14.8 ($4.24\cdotp10^{-34})^{**}$ $(5.01\cdotp 10^{-33})^{\dagger}$	& - & -	& 14.8 ($6.07\cdotp 10^{-34})^{**}$ $(7.29\cdotp 10^{-33})^{\dagger}$\\
& ta-2 & - & 13.0 ($1.98\cdotp 10^{-28})^{**}$ $(2.37\cdotp 10^{-27})^{\dagger}$ & -	& 19.9 ($2.74\cdotp 10^{-49})^{**}$ $(3.29\cdotp 10^{-48})^{\dagger}$\\
& ta-3 & -	& - & 17.0 ($1.10\cdotp 10^{-40}$) $(1.32\cdotp 10^{-39})^{\dagger}$	& 1.21 (0.22) (1) $\dagger$\\
\hline
\multirow{ 6}{*}{DTR} & at-1 & 20.10 ($1.02\cdotp 10^{-49})^{**}$ $(1.22\cdotp 10^{-48})^{\dagger}$	& - & -	& 33.44 ($1.18\cdotp 10^{-83})^{**}$ $(2.69\cdotp 10^{-82})^{\dagger}$\\
& at-2 & - & 33.9 ($2.06\cdotp 10^{-84})^{**}$ $(2.47\cdotp 10^{-83})^{\dagger}$ & - & 27.76 ($3.43\cdotp 10^{-70})^{**}$ $(4.12\cdotp 10^{-69})^{\dagger}$\\
& at-3 & -	& - & -0.57 (0.56) (0.15) $\dagger$ & 2.51 $(0.01)$ $(1)^{\dagger}$\\
& ta-1 & $2.49 (0.01)$ $(0.15)^{\dagger}$	& - & -	& 13.20 ($5.73\cdotp 10^{-29})^{**} $ $(6.88\cdotp 10^{-28})^{\dagger}$\\
& ta-2 & - & 11.6 ($2.75\cdotp 10^{-24})^{**} (3.30\cdotp 10^{-23})^{\dagger}$ & -	& 8.07 ($6.16\cdotp 10^{-14})^{**}$ $(7.93\cdotp 10^{-13})^{\dagger}$\\
& ta-3 & -	& - & 14.6 ($2.83\cdotp 10^{-33})^{**}$ $(3.40\cdotp 10^{-32})^{\dagger}$ & 11.21($6.54\cdotp10^{-23})^{**}$ $(7.85\cdotp 10^{-22})^{\dagger}$ \\
\hline
\end{tabular}
\end{table}
A comparison analysis of models average performance between original data and those enhanced with synthetic data is shown in table \ref{tab:table-5}.
\begin{table}[!ht]
\centering
\caption{\label{tab:table-5}Models performance increase across mental workload indexes between original dataset and dataset combined with synthetic data}
\vspace*{1mm}
\renewcommand{\arraystretch}{1}%
\begin{tabular}{c c p{0.9cm} p{0.9cm} p{0.9cm} p{0.9cm} p{0.9cm} p{0.9cm} p{0.9cm} p{0.9cm} p{0.9cm} p{1.3cm}} 
\hline
& Data & $\varB{c1-\theta}$ &	$\varB{c2-\theta}$ & $\varB{c3-\theta}$ & $\varB{c-\alpha}$ & $\varB{at-1}$ & $\varB{at-2}$ & $\varB{at-3}$ & $\varB{ta-1}$ &	$\varB{ta-2}$ & $\varB{ta-3}$ \\ [0.5ex] 
\hline\hline
\multirow{ 2}{*}{f1-score} & Orig. & 57.6\%	& 55.9\% & 58.8\% &	38.4\% & 53.8\% & 56.0\% & 44.1\% & 63.1\% & 56.4\% & 71.0\% \\
& Orig.+Synth. & 74.2\% & 64.6\% & 70.9\% &	62.8\% & 84.5\% & 82.2\% & 65.2\% & 70.3\% & 70.7\% & 60.8\% \\
\hline
\multirow{ 2}{*}{accuracy} & Orig. & 57.7\% & 56.6\% & 60.0\% &	50.0\% & 53.3\% & 56.6\% & 47.7\% & 63.3\% & 56.6\% & 71.0\% \\
& Orig.+Synth. & 74.8\% & 64.7\% & 70.0\% & 62.6\% & 84.5\% & 	82.2\% & 65.2\% & 70.3\% & 70.7\% &	60.8\% \\
\hline
\multirow{ 2}{*}{precision} & Orig. & 57.7\% & 56.7\% &	61.3\% &	33.3\% & 65.25\% & 57.2\% & 47.5\% &	63.8\% & 56.6\% & 71.1\% \\
& Orig.+Synth. & 74.2\% & 64.7\% & 70.6\% &	62.6\% & 84.6\% &	82.2\% & 65.2\% & 70.3\% & 70.7\% &	60.8\% \\  
\hline
\multirow{ 2}{*}{recall} & Orig. & 57.7\% &	56.6\% & 60.0\% &	50.0\% & 54.4\% & 56.6\% &	47.7\% & 63.3\%	& 56.6\% & 71.0\% \\
& Orig.+Synth. & 74.8\% & 64.5\% & 70.0\% &	62.0\% & 84.5\%	& 81.23\% & 65.2\% & 70.3\% &	70.7\% & 60.8\% \\ 
\hline
\multicolumn{2}{c}{\textbf{Avg. Increase:}} & \textbf{17.2}\%	& \textbf{8.0}\%	& \textbf{10.0}\% & \textbf{19.9}\% & \textbf{30.2}\% &	\textbf{25.5}\% & \textbf{18.4}\% &	\textbf{6.9}\% &	\textbf{14.1}\% & \textbf{-10.1}\% \\
\end{tabular}
\end{table}
An analysis of the number of electrodes across alpha and theta bands as given in table \ref{tab:table-1} outlined in the design section \ref{sec:design} we can see a higher number of electrodes in indexes $\varB{c1-\theta}$ and $\varB{c3-\theta}$ in comparison to indexes $\varB{c2-\theta}$ and $\varB{c-\alpha}$. To asses the impact of the number of electrodes in overall performance of the models, a cross-plotting between indexes $\varB{at-1}$ vs. $\varB{at-2}$ and $\varB{at-3}$ vs.$\varB{at-2}$ as well $\varB{ta-1}$ vs. $\varB{ta-2}$ and $\varB{ta-3}$ vs.$\varB{ta-2}$ is analyzed. Figure \ref{fig:6} illustrates this cross density plot comparison of performance between the alpha-to-theta and theta-to-alpha ratio indexes. 
Furthermore, a two-tailed significance test between these band ratio indexes ($\varB{at-1}$ and $\varB{at-3}$ versus $\varB{at-2}$ as well as $\varB{ta-1}$ and $\varB{ta-3}$ versus $\varB{ta-2}$) reveals a statistically significant difference. Table \ref{tab:table-6} presents the p-value significance levels for confidence interval of $\alpha = 0.05$. The p-values are with Bonferroni correction applied, resulting in a significance level set at $\alpha=0.005$.
\begin{table}[!ht]
\centering
\caption{\label{tab:table-6} The two-tailed t-test between alpha and theta band ratios: \textbf{$\varB{at-1}$} vs. \textbf{$\varB{at-2}$}: 2-tail test value between indexes $\varB{at-1}$ and $\varB{at-2}$. \textbf{$\varB{at-2}$} vs. \textbf{$\varB{at-3}$}: 2-tail test value between between indexes $\varB{at-2}$ and $\varB{at-3}$. \textbf{$\varB{ta-1}$} vs. \textbf{$\varB{ta-2}$}: 2-tail test value between indexes ta-1 and ta-2. \textbf{$\varB{ta-2}$} vs. \textbf{$\varB{ta-3}$}: 2-tail test value between indexes $\varB{ta-2}$ and $\varB{ta-3}$. The (\dag) sign indicate the p-value with Bonferroni correction applied, resulting in a significance level set at$\alpha=0.005$}
\vspace*{1mm}
\renewcommand{\arraystretch}{1}%
\begin{tabular}{c m{1.5cm} m{2.5cm} m{2.5cm} m{2.5cm} m{2.5cm}} 
\hline
 & Learning Technique & $\varB{at-1}$ vs. $\varB{at-2}$ & $\varB{at-3}$ vs. $\varB{at-2}$ & $\varB{ta-1}$ vs. $\varB{ta-2}$ & $\varB{ta-3}$ vs. $\varB{ta-2}$ \\ [0.5ex] 
\hline\hline
 \multirow{ 3}{*}{t-statistic (p-value)} & L-R & 4.82 ($2.79\cdotp10^{-6})^{**}$ ($1.11\cdotp10^{-5})^{\dagger}$  &-42.29 ($4.27\cdotp10^{-101})^{**}$ ($1.71\cdotp10^{-100})^{\dagger}$ &-4.98 ($1.32\cdotp10^{-6})^{**}$ ($5.29\cdotp10^{-6})^{\dagger}$  &-20.09 ($1.07\cdotp10^{-49})^{**}$ ($4.31\cdotp10^{-49})^{\dagger}$ \\ 
 & SVM & 3.31 ($1\cdotp10^{-2})^{**}$ ($4.34\cdotp10^{-3})^{\dagger}$ & -42.31 ($3.91\cdotp10^{-101})^{**}$ ($1.56\cdotp10^{-100})^{\dagger}$ & -4.10 ($6.00\cdotp10^{-5})^{**}$ ($2.40\cdotp10^{-4})^{\dagger}$ & -17.44 ($6.75\cdotp10^{-42})^{**}$ ($2.70\cdotp10^{-41})^{\dagger}$ \\ 
 & DTR & 6.96 ($4.76\cdotp10^{-11})^{**}$ ($1.90\cdotp10^{-10})^{\dagger}$ & -29.21 $(9.99\cdotp10^{-74})^{**}$ $(3.93\cdotp10^{-73})^{\dagger}$ & 5.73 $(3.64\cdotp10^{-8})^{**}$ $(1.45\cdotp10^{-7})^{\dagger}$ & -20.99 $(2.93\cdotp10^{-52})^{**}$ $(1.17\cdotp10^{-51})^{\dagger}$\\ 
 \hline
\end{tabular}
\end{table}

\section{Discussion}
\label{sec:conclusions}
Rapid advancements in various tools and technologies introduced new perspectives in using EEG signals to classify task load conditions using machine learning techniques. The analysis done so far on EEG frequency bands, specifically alpha and theta bands, seems to correlate the changes in these bands to task load \citep{gevins2003neurophysiological, borghini2014measuring}. Researchers face many problems in using EEG band ratios for the purpose of mental workload modelling:  i) the limited amount of participants for each conducted empirical experiment ii) a clear definition of mental workload iii) a clear EEG measure of mental workload. 

In detail, the three aforementioned issues can be overcome and this article is a testament for such a claim. In fact, this research work demonstrates how the first issue can be tackled by using modern deep-learning methods for synthetic data generation, giving the possibility to expand the often limited cardinality of existing datasets created with EEG data. It also contributes to tackle the second issue by advancing the understanding of mental workload as a construct by means of an empirical experiment with EEG data. In particular, it performs a construction of indexes of mental workload by employing the alpha and theta EEG bands individually and in combination, and the extraction of statistical features from these indexes for the discrimination of self-reported perceptions of mental workload.

Results show that, from an initial highest accuracy of $60\%$ for the individual alpha and theta indexes on the original dataset, we witnessed an increase between $8\%$ to $20\%$ in classifier performance when this data has been augmented with synthetic data. 

Regarding mental workload ratio indexes, especially the alpha-to-theta indexes, it was possible to build models with minimum $18.4\%-30.2\%$ higher performance (as measured by f1-score, accuracy, precision, recall) than the other indexes. Furthermore, the results show that mental workload indexes $\varB{at-1}$, $\varB{at-2}$ and $\varB{ta-1}$, $\varB{ta-2}$ can better discriminate self-reported perceptions of mental workload in comparison to their individual counterparts ($\varB{c-\alpha}$, $\varB{c1-\theta}$ and $\varB{c2-\theta}$). This proves our hypothesis \textbf{H1} given earlier that alpha-to-theta and theta-to-alpha ratios can significantly discriminate self-reported perceptions of mental workload than the individual use of EEG band power and can be used in designing highly accurate classification models. The accuracy, f1-score, recall and precision evaluation metrics indicate a good classification across almost all alpha-to-theta and theta-to-alpha indexes. 

One interesting observation is the impact of the number of electrodes in the selected indexes on the overall accuracy of the classifiers. For example, it can be seen from table \ref{tab:table-1} that $\varB{c1-t}$ from theta band has a higher number of electrodes that contribute to the computation of band ratios and indicate the higher accuracy in both alpha-to-theta and theta-to-alpha indexes. 
Given the results from figure \ref{fig:6} and \ref{tab:table-6}, hypothesis \textbf{H2} cannot be conclusively proven that the number of electrodes used for calculating alpha-to-theta and theta-to-alpha better effectuate the predictive power of the classifiers. Figure \ref{fig:6} indicates better performance of $\varB{at-2}$ and $\varB{ta-2}$ indexes which are computed from $\varB{c2-\theta}$ and $\varB{c-\alpha}$ individual indexes that, if seen from table \ref{tab:table-1} have lesser electrode numbers. One potential explanation hypothesised by authors lies in the nature of the experiment performed while collecting STEW datasets' EEG recordings, where ``Rest'' and ``Simkap'' activities are performed in sequence one after the other. Some research indicates a strong correlation between EEG frequency patterns and the relative levels of distinct neuromodulators \citep{vakalopoulos2014eeg}. This sudden change in task load activity may lead to neuromodulation in the parietal region and neuronal suppression on the frontal cortical region, resulting in better performance of band ratio indexes ($\varB{at-2}$ and $\varB{ta-2}$) with a smaller number of electrodes. Further research is required to validate this claim.

Based on the results above, we can conclude that EEG band ratios, alpha-to-theta and theta-to-alpha ratio mental workload indexes, can significantly discriminate self-reported perceptions of mental workload and be used to design models for detecting such levels of mental workload perception. The observations however cannot conclusively prove that the higher the electrode number, especially in the parietal region, leads to a better discrimination self-reported perceptions of mental workload.

\section{Conclusion}
Various EEG frequency bands indicate a direct correlation to human mental workload. In particular, EEG bands such as alpha and theta bands tend to increase/decrease in the state of mental workload \citep{borghini2014measuring}. However, no conjoint analysis of both bands in the form of indexes over time has been sufficiently analysed so far. 

This article has empirically demonstrated that EEG band ratios, specifically the alpha-to-theta and theta-to-alpha ratios can be treated as mental workload indexes for the discrimination of self-reported perceptions of mental workload. In detail, a set of higher level features associated to these indexes, have proven useful for the inductive formation of models, employing machine learning, for the discrimination of two levels of mental workload perception (``suboptimal MWL'' and ``super optimal MWL''). Another important contribution in this research is the analysis of the impact of electrode density in-band ratios on the formation of discriminative models of self-reported perceptions of mental workload.

Future research work will outline the usage of the alpha-to-theta and theta-to-alpha ratio indexes related to the following issues:

\begin{itemize}
    \item replication of the experiment conducted in this research with additional public available datasets, to further validate the contribution to knowledge.
    \item evaluation of human tasks different than those employed in this research, as for instance those conducted in the automobile industry \citep{di2018eeg}, in the context of Human-Computer Interaction (HCI) \citep{longo2012formalising} and in education \citep{longo2018reliability, longo2018evaluation}.
    \item use of multi-channel EEG data collected from a larger pool of electrodes, and thus formation and evaluation of additional mental workload indexes built with different clusters of electrodes for the alpha and theta bands.
    \item the design of a novel experiment with additional task load conditions of incremental complexity, for example by employing the multiple resource theory of Wickens \citep{wickens2008multiple}  and the definition of objective task performance measures that can be used as dependent features from indexes of mental workload.
\end{itemize}

\section*{Conflict of Interest Statement}

The authors declare that the research was conducted in the absence of any commercial or financial relationships that could be construed as a potential conflict of interest.

\section*{Author Contributions}
Longo L., and Raufi B. designed the study. Bujar, R. conducted the experiment. All authors reviewed and approved the final manuscript.

\section*{Funding}
The research is part of MCSA Post-doc CareerFIT fellowship, funded by Enterprise Ireland and the European Commission. Fellowship ref. number: MF2020 0144

\section*{Data Availability Statement}
The STEW datasets analyzed for this study can be found in STEW: Simultaneous Task EEG Workload Dataset - https://ieee-dataport.org/open-access/stew-simultaneous-task-eeg-workload-dataset.

\bibliographystyle{frontiersinSCNS_ENG_HUMS} 
\bibliography{frontiers}


\end{document}


\onecolumn
\firstpage{1}

\title[Supplementary Material]{{\helveticaitalic{Supplementary Material}}}

\maketitle

\section{Supplementary Tables and Figures}

\subsection{Tables}

\begin{longtable}{cc} 
\caption{List of extracted features.}
\label{s-table2}
No & Feature Name \\
\hline
1...76 & FFT mean coefficient\_\{0...75\}\\
77 & Fundamental frequency\\
78 & Human range energy \\
79...91 & LPCC\_\{0...12\} \\
92...103 & MFCC\_\{0...11\} \\
104 & Maximum power spectrum \\
105 & Maximum frequency \\
106 & Median frequency \\
107	& Power bandwidth \\
108	& Spectral centroid \\
109	& Spectral decrease \\
110	& Spectral distance \\
111	& Spectral entropy \\
112	& Spectral kurtosis \\
113	& Spectral positive turning points \\
114	& Spectral roll-off\\
115	& Spectral roll-on\\
116	& Spectral skewness\\
117	& Spectral slope\\
118	& Spectral spread\\
119	& Spectral variation\\
120...128	& Wavelet absolute mean\_\{0...8\}\\
129...137	& Wavelet energy\_\{0...8\}\\
138	& Wavelet entropy\\
139...147	& Wavelet standard deviation\_\{0...8\}\\
148...156	& Wavelet variance\_\{0...8\}\\
157...166	& ECDF\_\{0...9\}\\
167, 168	& ECDF Percentile\_\{0, 1\}\\
169, 170	& ECDF Percentile Count\_\{0, 1\}\\
171...180	& Histogram\_\{0...9\}\\
181	& Interquartile range\\
182	& Kurtosis\\
183	& Max\\
184	& Mean\\
185	& Mean absolute deviation\\
186	& Median\\
187	& Median absolute deviation\\
188	& Min\\
189	& Root mean square\\
190	& Skewness\\
191	& Standard deviation\\
192	& Variance\\
193	& Absolute energy\\
194	& Area under the curve\\
195	& Autocorrelation\\
196	& Centroid\\
197	& Entropy\\
198	& Mean absolute diff\\
199	& Mean diff\\
200	& Median absolute diff\\
201	& Median diff\\
202	& Negative turning points\\
203	& Neighbourhood peaks\\
204	& Peak to peak distance\\
205	& Positive turning points\\
206	& Signal distance\\
207	& Slope\\
208	& Sum absolute diff\\
209	& Total energy\\
210	& Zero crossing rate
\end{longtable}

\subsection{Figures}


\begin{figure}[!ht]
    \includegraphics[width=9cm]{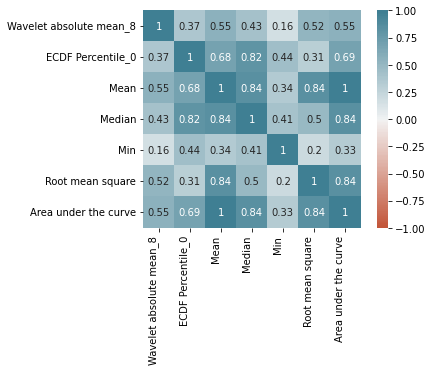}
    \includegraphics[width=9cm]{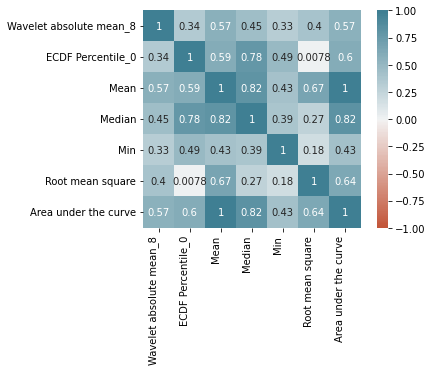}
    \caption{Correlation Matrix for the Case - \textbf{c1-t} - Rest (left) and Simkap (Right). \textbf{c1-t}: index of theta cluster \#1.}
\end{figure}
\begin{figure}[!ht]
    \includegraphics[width=9cm]{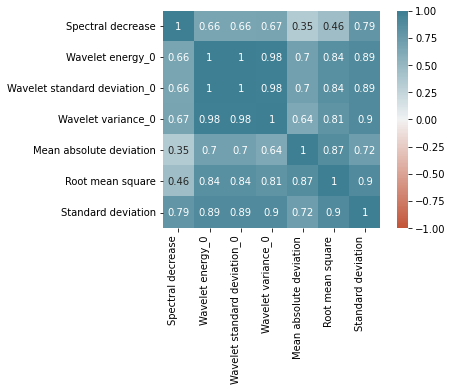}
    \includegraphics[width=9cm]{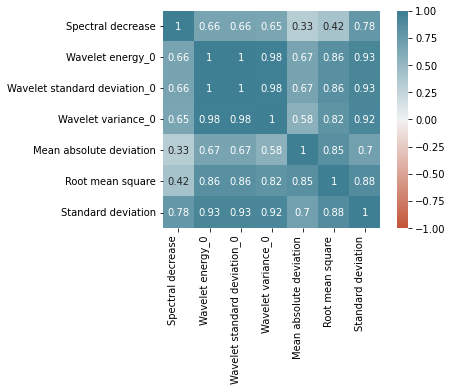}
    \caption{Correlation Matrix for the Case - \textbf{c1-t} - Rest (left) and Simkap (Right). \textbf{c2-t}: index of theta cluster \#2.}
\end{figure}
\begin{figure}[!ht]
    \includegraphics[width=9cm]{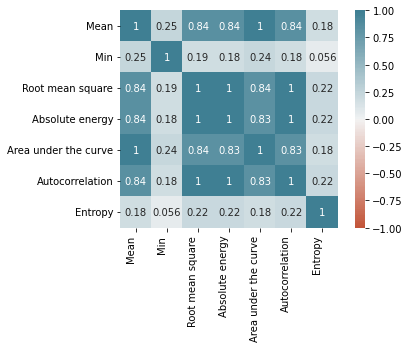}
    \includegraphics[width=9cm]{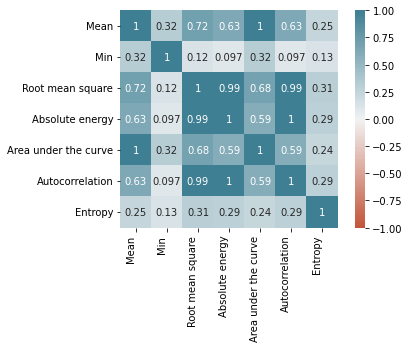}
    \caption{Correlation Matrix for the Case - \textbf{c3-t} - Rest (left) and Simkap (Right). \textbf{c3-t}: index of theta cluster \#3.}
\end{figure}
\begin{figure}[!ht]
    \includegraphics[width=9cm]{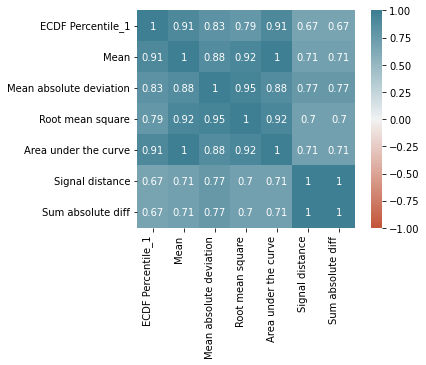}
    \includegraphics[width=9cm]{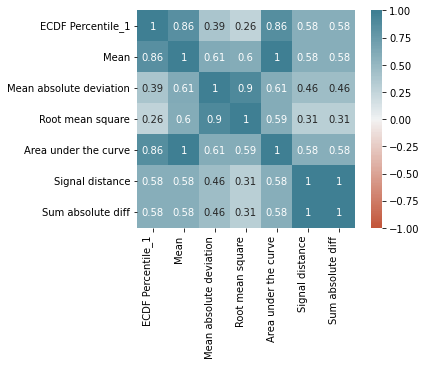}
    \caption{Correlation Matrix for the Case - \textbf{c-a} - Rest (left) and Simkap (Right). \textbf{c-a}: index of alpha cluster.}
\end{figure}
\begin{figure}[!ht]
    \includegraphics[width=9cm]{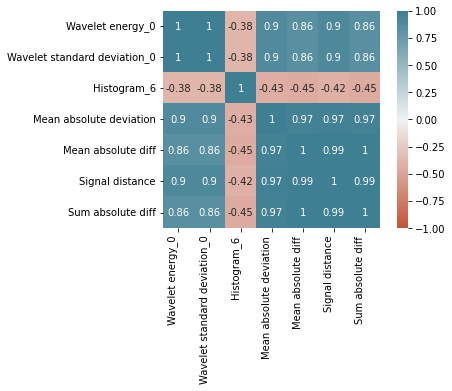}
    \includegraphics[width=9cm]{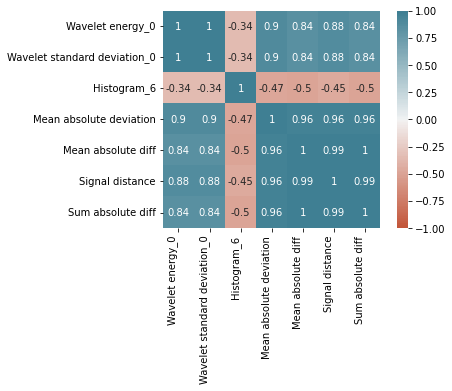}
    \caption{Correlation Matrix for the Case - \textbf{at-2} - Rest (left) and Simkap (Right). \textbf{at-2}: alpha-to-theta ratios between the cluster of alpha and cluster 2 of theta band.}
\end{figure}
\begin{figure}[!ht]
    \includegraphics[width=9cm]{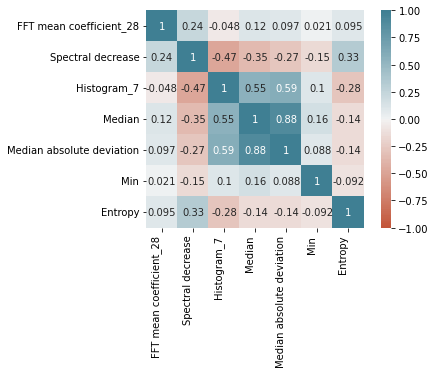}
    \includegraphics[width=9cm]{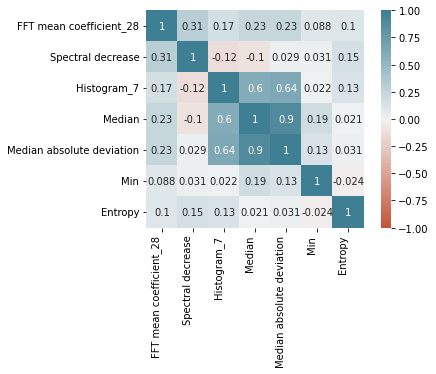}
    \caption{Correlation Matrix for the Case - \textbf{at-3} - Rest (left) and Simkap (Right). \textbf{at-3}: alpha-to-theta ratios between the cluster of alpha and cluster 3 of theta band.}
\end{figure}
\begin{figure}[!ht]
    \includegraphics[width=9cm]{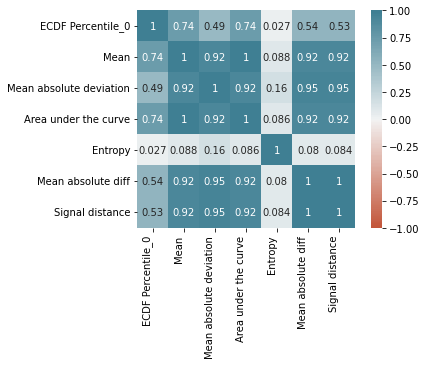}
    \includegraphics[width=9cm]{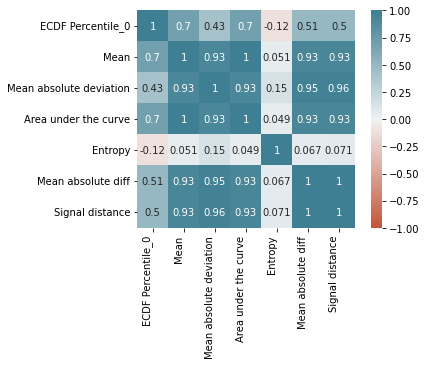}
    \caption{Correlation Matrix for the Case - \textbf{ta-1} - Rest (left) and Simkap (Right). \textbf{ta-1}: theta-to-alpha ratios between the cluster 1 of theta and cluster of alpha band.}
\end{figure}
\begin{figure}[!ht]
    \includegraphics[width=9cm]{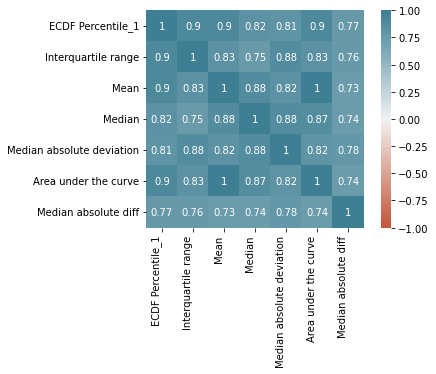}
    \includegraphics[width=9cm]{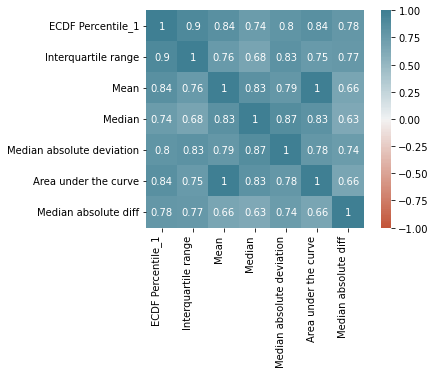}
    \caption{Correlation Matrix for the Case - \textbf{ta-2} - Rest (left) and Simkap (Right). \textbf{ta-2}: theta-to-alpha ratios between the cluster 2 of theta and cluster of alpha band.}
\end{figure}
\begin{figure}[!ht]
    \includegraphics[width=9cm]{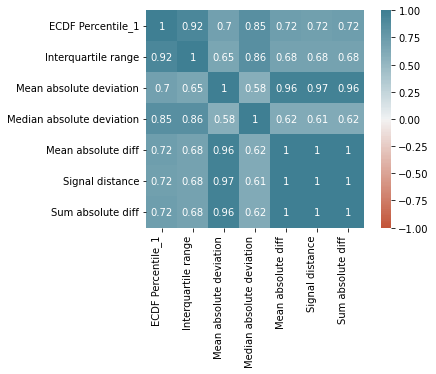}
    \includegraphics[width=9cm]{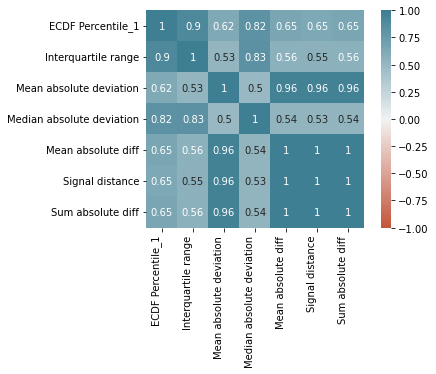}
    \caption{Correlation Matrix for the Case - \textbf{ta-3} - Rest (left) and Simkap (Right). \textbf{ta-3}: theta-to-alpha ratios between the cluster 3 of theta and cluster of alpha band.}
\end{figure}

\begin{figure}[h!]
    \includegraphics[width=18cm]{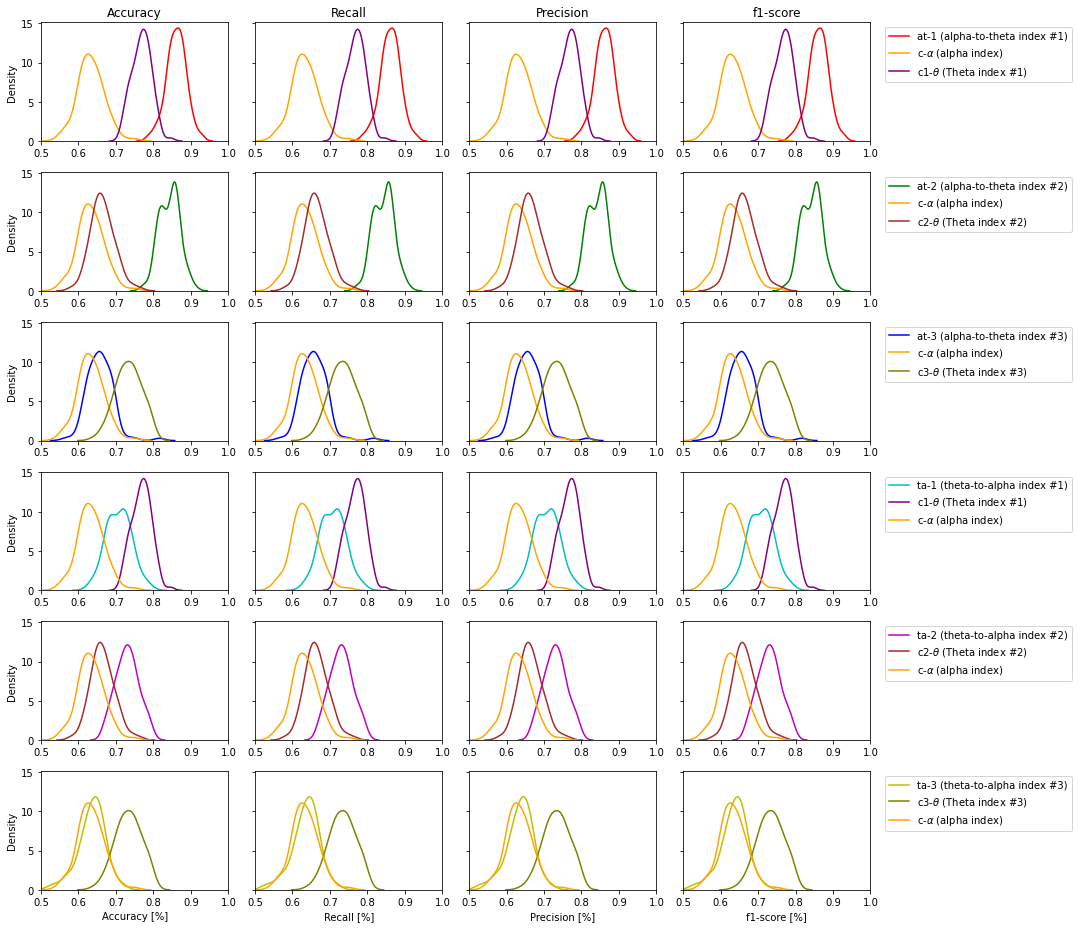}
    \caption{Density plots for Logistic Regression (L-R) across all performance metrics between band ratio indexes vs. their individual indexes}
\end{figure}

\begin{figure}[h!]
    \includegraphics[width=18cm]{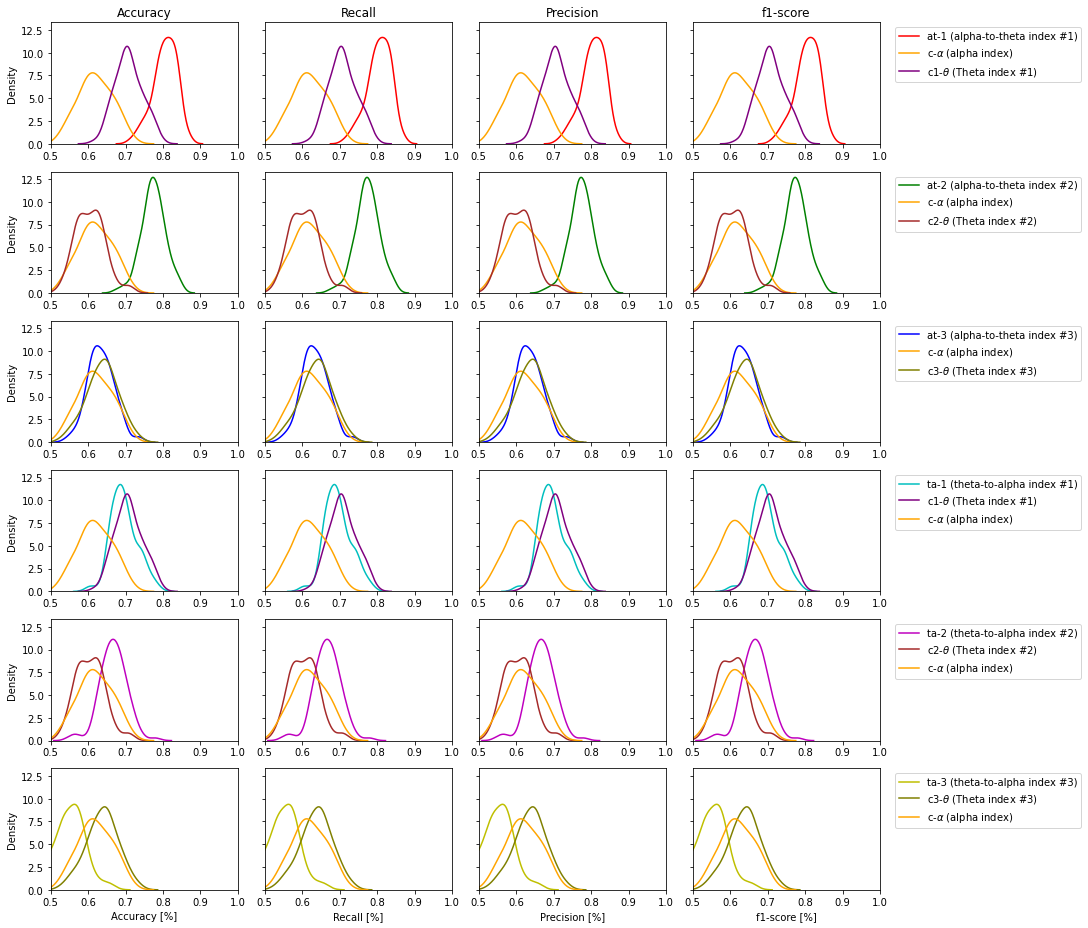}
    \caption{Density plots for Decision Tree (DTR) across all performance metrics between band ratio indexes vs. their individual indexes}
\end{figure}

\begin{figure}[ht!]
    \includegraphics[width=18cm]{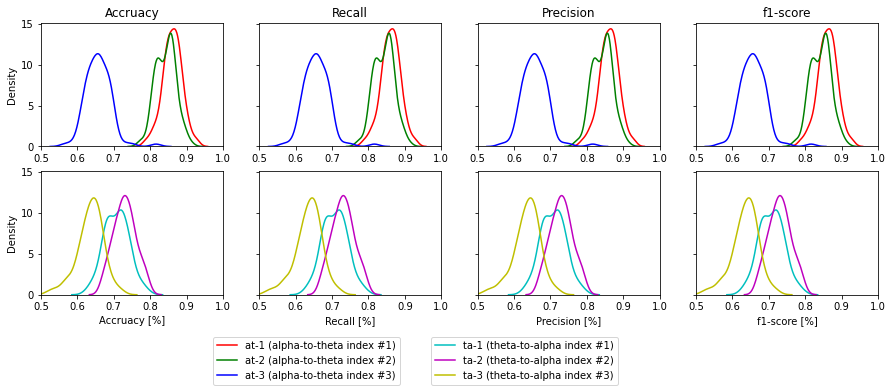}
    \caption{Density plots across all performance metrics between all band ratio indexes (the case for Logistic Regression (L-R))}
    \label{fig:6}
\end{figure}

\begin{figure}[ht!]
    \includegraphics[width=18cm]{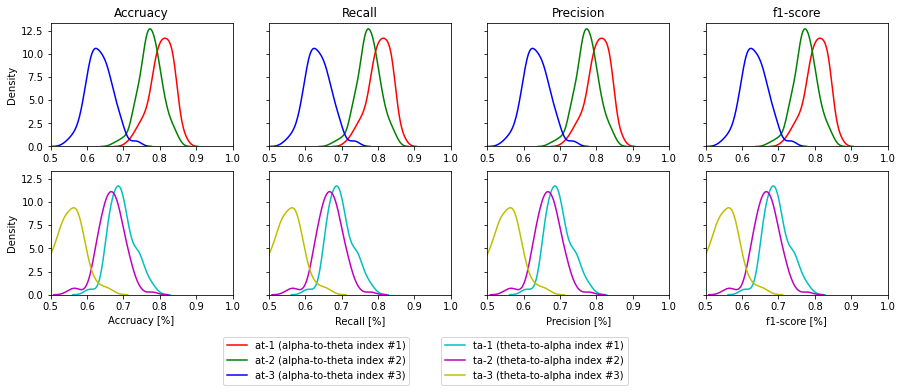}
    \caption{Density plots across all performance metrics between all band ratio indexes (the case for Decision Trees (DTR))}
    \label{fig:6}
\end{figure}

